\documentclass[lettersize,journal]{IEEEtran}
\usepackage{amsmath,amsfonts}
\usepackage{algorithmic}
\usepackage{algorithm}
\usepackage{array}
\usepackage[caption=false,font=normalsize,labelfont=sf,textfont=sf]{subfig}
\usepackage{textcomp}
\usepackage{stfloats}
\usepackage{url}
\usepackage{verbatim}
\usepackage{graphicx}
\usepackage{cite}
\usepackage{multicol}
\usepackage{multirow}
\usepackage{graphics} % for pdf, bitmapped graphics files
\usepackage{longtable}
\usepackage{rotating}
\usepackage{booktabs}
\usepackage{makecell}
\usepackage{setspace}

\begin{document}

\title{Recent Developments in Recommender Systems: A Survey}

\author{Yang Li,~\IEEEmembership{Member,~IEEE},~Kangbo Liu,~Ranjan Satapathy,~Suhang Wang,~\IEEEmembership{Member,~IEEE},

~and Erik Cambria,~\IEEEmembership{Fellow,~IEEE}
        % <-this % stops a space
\thanks{This paper was produced by the IEEE Publication Technology Group. They are in Piscataway, NJ.}% <-this % stops a space
\thanks{Manuscript received April 19, 2021; revised August 16, 2021.}}

\markboth{Journal of \LaTeX\ Class Files,~Vol.~14, No.~8, August~2021}%
{Shell \MakeLowercase{\textit{et al.}}: A Sample Article Using IEEEtran.cls for IEEE Journals}


\maketitle

\begin{abstract}
In this technical survey, we comprehensively summarize the latest advancements in the field of recommender systems. The objective of this study is to provide an overview of the current state-of-the-art in the field and highlight the latest trends in the development of recommender systems. The study starts with a comprehensive summary of the main taxonomy of recommender systems, including personalized and group recommender systems, and then delves into the category of knowledge-based recommender systems. In addition, the survey analyzes the robustness, data bias, and fairness issues in recommender systems, summarizing the evaluation metrics used to assess the performance of these systems. Finally, the study provides insights into the latest trends in the development of recommender systems and highlights the new directions for future research in the field. 
\end{abstract}


\section{Introduction}
\IEEEPARstart{T}{he} utilization of recommender systems is a crucial aspect in addressing the issue of online information overload and improving customer relationship management. These systems are designed to provide personalized recommendations to users of online products and services, enhancing the user's online experience. The application of recommender systems can be seen in various online platforms, such as playlist generators for listeners, product recommendations for customers, and content recommendations for readers, among others. The objective of these systems is to identify new and relevant items that align with the user's preferences, leading to an increase in incremental revenue for companies. The fundamental principle of recommender systems is to suggest relevant items to users by leveraging feature engineering techniques on user preferences, item features, and their interactions (such as purchases or clicks).
The categorization of recommender systems can be based on the number of recommended users, which can be classified into two main types: personalized and group recommender systems. The personalized recommender system is designed to predict the next item for an individual user based on their past preferences and behaviors, while the group recommender system considers the collective preferences of a group of users to provide a recommendation that alleviates conflicts within the group. The personalized recommender system is the most widely researched and studied type, with a focus on personalizing the recommendations to the individual user's interests. In the past decade, previous studies have categorized personalized recommender systems into three main approaches: collaborative filtering, content-based, and hybrid methods.
A collaborative filtering-based recommender system is one of the most widely used approaches and it makes recommendations based on the behavior of other users. This type of recommender system can be broadly classified into three categories: memory-based, model-based, and context-aware based. In the context-aware based category, researchers focus on incorporating additional information such as user preferences, emotions, and item attributes in the model design. This has led to an increase in the use of deep learning for feature selection and feature addition.
The content-based recommender system is another popular approach that classifies items based on their attribute information and provides recommendations accordingly. This method eliminates the cold start and sparse problems and has the advantage of providing explanations for the recommended items by listing the content features that caused the recommendation.
Recently, the use of knowledge extracted from user profiles or public data has become more prevalent in solving problems such as the cold start and grey sheep problems~\cite{angbri} (The term ``grey sheep'' refers to items or users that do not fit into the traditional categories or clusters in a recommendation system, resulting in suboptimal recommendations). This has resulted in the extension of the categories to include collaborative filtering-based, content-based, knowledge-based, and hybrid methods.

Whereas, group recommendation will be more difficult as it needs to consider more than one people's interests. 
Based on the literature in recent years, we divide them into two categories, i.e., memory-based and model-based. 
However, we cannot include all types of recommender systems (e.g., multi-criteria recommender system, risk-aware recommender system, mobile recommender system, etc.). To make a mainstream summarization, we mainly focus on the most popular category in the group recommender system as we have described above.

As a practical commercial application, the robustness, data bias, and fairness problem of a recommender system play crucial roles in its commercial viability.
Robustness is the ability of a recommender system to maintain accurate and effective recommendations in the face of varying or unexpected data.
Data bias is the systematic deviations in the data used by a recommender system that can result in unfair or inaccurate recommendations.
Fairness denotes the principle that a recommender system should provide unbiased and equitable recommendations for all users, regardless of their demographic or personal characteristics.

In terms of evaluation metrics, recommender systems can be classified into Rating Based Indicators (RBI) and Item Based Indicators (IBI). RBI evaluates the recommendations based on a predicted rating score, while IBI evaluates the recommendations based on a set or list of predicted items. This taxonomy allows for the classification of most existing evaluation indicators.

Through this summary,  we aim to provide a comprehensive overview of the current state of recommender systems, including their categorization, implementation, and evaluation metrics.

\textbf{Difference between previous surveys:}
There have been several survey papers on recommender systems in the past, but they have focused only on a specific domain, such as contextual aware-based recommender systems. For instance, Haruna et al.~\cite{haruna2017context} reviewed 178 technique papers published between January 2010 and October 2017 and presented state-of-the-art results and applications in this domain. Similarly, survey works by Raza et al.~\cite{raza2019progress} and Abdulkarem et al.~\cite{abdulkarem2019context} provided an overview of the current trends and techniques in contextual aware-based recommender systems.

Recently, deep learning and knowledge graphs have had a significant impact on recommender systems, leading to many works in these areas. Zhang et al.~\cite{zhang2019deep} focused on deep learning in recommender systems, while Guo et al.~\cite{guo2020survey} summarized knowledge graph-based recommender systems. Gao et al.~\cite{gao2020deep} and Wu et al.~\cite{wu2022graph} provided overviews of deep learning and graph neural networks in recommender systems, respectively.

Our survey, however, provides a comprehensive overview of recommender systems, covering the problems and new techniques in the field. We have created a new taxonomy and discussed common evaluation metrics and challenges, such as robustness, data bias, and fairness issues. Our survey stands out from others as it offers a broader perspective and covers more aspects of recommender systems.

\section{Personalized Recommender Systems}
\label{sec:personalizedRS}
With the advent of deep learning and the abundance of data available online, personalized recommender systems have become increasingly prevalent in our lives, making it easier and more efficient for people to experience new things~\cite{zhang2019deep}. These systems are commonly categorized into three main types: Collaborative Filtering based, Context-aware based, and Hybrid. However, recent developments have seen the integration of external knowledge into these systems, leading to the creation of knowledge-based recommender systems. To further expand upon this, the paper will present an in-depth examination of knowledge-based recommender systems in section~\ref{sec:kbs}.
\subsection{Collaborative Filtering Based RS}
\label{subsec:colabrs}
Collaborative filtering (CF) is a popular technique used in recommendation systems to predict a user's preferences or opinions by leveraging the collective knowledge of a large pool of users. CF operates based on the idea that users who share similar preferences and behaviors will also have similar opinions. This method of recommendation is driven by the user profiles that are created from their historical interactions, such as their past browsing history, the items they have purchased, or the songs they have listened to. The CF approach can be divided into two main categories: memory-based CF and model-based CF. Memory-based CF utilizes the users' profiles to make recommendations in a simpler, more direct way. On the other hand, model-based CF employs more sophisticated mathematical models to generate recommendations, which often result in more accurate and reliable predictions.

\subsubsection{Memory-based CF}
Memory-based CF is a widely used recommendation technique that leverages the similarity between users or items to make recommendations. The algorithm is comprised of two key steps: first, the similarity between users or items is calculated using a nearest neighbor-based method; and second, recommendations are generated for users based on the sorted similarity values. This approach is simple and efficient, making it a popular choice for many recommender systems. The ability to capture user behavior patterns and preferences based on historical data, as well as its ability to handle large amounts of data, makes memory-based CF a valuable tool for personalized recommendation.

However, despite the simplicity and popularity of memory-based CF, it often encounters challenges in the form of sparsity in the interaction matrix and computational difficulties in large user or item spaces. The sparsity of the interaction matrix implies that there are only a limited number of non-zero values, which makes the prediction of recommendations more susceptible to errors. Moreover, as the size of the user and item spaces grows, the calculation of similarity becomes increasingly complex and time-consuming, making it difficult to scale up this method in large-scale applications.
The calculation of similarity between users or items is a crucial aspect of recommender systems. In collaborative filtering, the similarity is primarily obtained from the co-rated items and is used to predict the preference of an item or user. There are various methods to calculate the similarity, such as Pearson Correlation Coefficient (PCC), Cosine similarity (COS), Mean Squared Difference (MSD), Proximity Impact Popularity (PIP), Jaccard Similarity, and Proximity Significance Singularity (PSS). Each of these methods has its own strengths and limitations, and the selection of a specific method depends on the type of recommender system and the data used. In this paper, we will provide a comprehensive overview of these similarity calculation methods in subsection~\ref{subsec:sim}.
To reduce the computational complexity in recommender systems, the use of embedding techniques has been widely adopted by researchers. The aim of these techniques is to map high-dimensional sparse vectors to low-dimensional dense vectors and calculate the similarity between users or items. For instance, Valcarece et al. proposed the ``prefs2vec'' model, which is inspired by the bag-of-words (CBOW) in ``word2vec'' to learn the user and item embedding~\cite{valcarce2019collaborative}. On the other hand, Chen et al. applied a graph-based method to learn the user and item representation~\cite{chen2019collaborative}. These approaches have proven to be effective in reducing computational complexity, making the recommender system more efficient and faster. 
Barkan et al.~\cite{barkan2021anchor} introduced an anchor vector-based recommender system. This paper applies an inductive variant of a conformal recommender system and proposed nonconformity measures in the inductive setting.

Memory-based CF is one of the most classic methods in recommendation systems. It calculates the similarity between users or items based on their past behaviors, such as rating scores or click data, and generates recommendations by finding the nearest neighbors. The strength of memory-based CF lies in its simplicity and high accuracy in small-scale scenarios. However, its scalability becomes a challenge when dealing with large-scale data due to the huge computational cost of similarity calculation. Therefore, many researchers have tried to improve the efficiency of memory-based CF by adopting various techniques, such as dimensionality reduction, sampling, and parallel computing. In summary, memory-based CF is a fundamental and effective method in recommendation systems, but its performance needs to be carefully considered in practical applications.

\subsubsection{Model-based CF}

Model-based CF is another widely adopted recommendation method that predicts the preferences of users based on the relationships between them and the items. 
In recent years, there has been a growing interest in the application of neural networks in recommendation systems. The integration of review text data has been shown to improve the performance of recommendation systems. This is why models like Convolutional Neural Network (CNN), Long Short-Term Memory (LSTM), Recurrent Neural Network (RNN), and AutoEncoder have been widely adopted in recent studies~\cite{choudhary2023sarwas,an2022design,liu2022review}, and they can be named neural-based CF in general. One of the approaches is utilizing Multi-Layer Perceptron (MLP) for latent feature learning~\cite{li2022mlp4rec}, resulting in successful models such as Neural Collaborative Filtering (NCF)~\cite{he2017neural} and Neural Network Matrix Factorization (NNMF)~\cite{dziugaite2015neural}. These models provide personalized recommendations to users and have proven to be highly effective~\cite{chen2022survey}.
The basic framework of the neural-based CF consists of mapping user and item representations, $s_{u}$ and $s_{i}$, respectively, as well as reviewing text data, $t_{ui}$, into a continuous vector space through a neural network $NN$. 

\begin{equation}
\small
\hat{r} = NN(s_{u},s_{i}, t_{ui}\vert\theta)
\end{equation}
The model parameters $\theta$ are then optimized to minimize the prediction error. The output of the neural network, $\hat{r}$, represents the predicted rating and is usually fed into a factorization-based method for further refinement.

The Factorization Machine (FM) is a powerful tool in the field of neural-based CF. It offers the ability to handle sparse data efficiently, which is a common challenge in user-item interaction data. This makes it ideal for use in real-world recommendation tasks, as it can incorporate a variety of features, including both categorical and numerical features. Its scalability in training procedures also makes it a popular choice for large-scale recommendation systems. FM models user-item interactions effectively, making it an indispensable component in many recommender systems.
The factorization machine predicts the output $y\in R$ based on equation~\ref{equ:fm}, which considers the dot product between $w$ and $x$ as well as the dot product between the rows of $V$. The weight parameter $w\in R^{d}$ and the factorization matrix $V\in R^{d\times k}$ are key components of the model, with $k$ being a rank hyper-parameter that can be adjusted based on the data type. For instance, if the data is text data, $k$ represents the vocabulary size. The computational complexity of the factorization machine is $O(dk)$, making it faster than traditional models with complexity $O(d^{2})$. It is worth noting that matrix factorization still has great potential, Kawakami et al. \cite{kawakami2021investigating} achieved performance comparable to other techniques by hyper parameterizing non-negative matrix factorization.

\begin{equation}
	\label{equ:fm}
 \small
	y_{FM} := <w,x> + \sum_{i>j}<\overline{v}_{j}, \overline{v}_{i}> x_{j}x_{i}
\end{equation}

Plenty of works apply this method to construct their models. For example,
Zheng et al.~\cite{zheng2017joint} adopted two parallel CNNs to extract features from user review text and item review text respectively, and constructed the model DeepCoNN to link user behaviors and item properties together, and fed the learned features into a factorization machine to construct the loss function. 
Similarly, Seo et al.~\cite{seo2017interpretable} model the user and the item separately with CNN based model. The difference is that in this work, 
\begin{itemize}
	\item they applied the local and global attention mechanism to reduce the noise, and
	\item they simply used the inner product of the user and item representation for the rating prediction.
\end{itemize}
The above works mainly extract features at the sentence level. While Chin et al.~\cite{chin2018anr} argued that not all parts of the reviews are equally important, they proposed an aspect-based neural recommender (ANR) to learn more granular feature representations of items. Li et al.~\cite{li2019capsule} applied the capsule neural network to extract the viewpoints and aspects from the user and item review text, respectively.
Wu et al.~\cite{wu2019npa} used CNN to be \textit{news encoder} and \textit{user encoder} which are to learn news representation and user representation based on his/her clicked news respectively, based on these two encoders, the personalized attention is applied to boost the performance. 
As we know, the explanation is an important factor in improving user experience improving and discovering system defects. Liu et al.~\cite{liu2020explainable} built the explainable recommender system by resolving the learning representation with a graph neural network.
Fang et al.~\cite{fang2022differentially} applied the variational autoencoder model to provide a precise recommendation based on sensitive information.

Interestingly, due to the complexity of the relationship between users and items, together with Spectral CNN~\cite{bruna2013spectral}, GCN~\cite{kipf2016semi}, GraphSAGE~\cite{hamilton2017inductive}, GAT~\cite{velickovic2017graph} and other classic graph neural network proposals the graph neural network model has been widely used in the field of recommendation.JODIE\cite{kumar2019predicting} model in the field of music recommendation, NGCF~\cite{wang2019neural}, Multi-GCCF~\cite{sun2019multi},~DGCF\cite{wang2020disentangled}, LightGCN~\cite{he2020lightgcn} in POI(Point of Interest) recommendation field and book recommendation field have achieved good results, among which LightGCN has become a classic model in the RS field.

In recent years, Wu et al.~\cite{ wu2021self} improved the robustness and accuracy of the graph neural network through three operations: node discarding, edge discarding and random walk. Xia et al.~\cite{xia2021incremental} proposed an incremental graph convolutional network (IGCN) model including meta-learning to solve the ``catastrophic forgetting'' problem and cold start problem of GNN. Lin et al.~\cite{ lin2022improving} introduced comparative learning based on neighbor nodes on the basis of a graph neural network, which greatly improved the performance of the model. Li et al.~\cite{li2022hypercomplex} considered the recommendation behavior in the hypercomplex number space, and designed a hypercomplex graph convolution operator based on the Cayley–Dickson structure, which improved the performance of the model. Xia et al.~\cite{xia2022hypergraph} used hypergraphs to enhance the capture ability of the cross-view contrastive learning architecture, which enhanced the representation ability of the model and improved the performance and robustness of the model. Zhao et al.~\cite{zhao2022investigating} proposed the r-AdjNorm plug-in that controls the normalization strength of neighborhood aggregation, which can solve the problem of popularity bias caused by graph neural networks.

As shown in the table~\ref{tab:graph_ndcg_recall1}, we summarize the performance of the RS model based on the graph neural network under the MovieLens dataset\footnote{https://grouplens.org/datasets/movielens/ }. We choose NDCG@K(Equation~\ref{equ:ndcg}) and Recall@K(Equation~\ref{equ:compre})  as the evaluation criteria, where K represents the system's TOP-K recommendation. In order to unify the data, we unify the model results to two decimal places according to the principle of rounding. From the table, we observe seven methods, consisting of five basic models and two combined models, that belong to three categories: graph collaborative filtering, graph convolutional neural networks, and self-supervised graph learning. Notably, the two combined models \cite{xia2021incremental, lin2022improving}, which are based on the collaborative filtering algorithm of graph neural networks, achieve the best results. The neighborhood-enhanced contrastive learning method, integrated into the collaborative filtering algorithm, can alleviate the ``catastrophic forgetting'' issue and avoids cold starts, producing slightly better outcomes than the combined incremental graph convolution methods \cite{lin2022improving}. As for the basic models of graph collaborative filtering, results of the three models \cite{wang2019neural, sun2019multi, wang2020disentangled} are comparable, and \cite{wang2020disentangled} slightly outperforms the other two. Finally, the basic models of graph convolutional neural networks \cite{he2020lightgcn} and self-supervised graph learning \cite{he2020lightgcn} exhibit similar performance.

\begin{table}[]
\centering
\label{tab:graph_ndcg_recall1}
\caption{Performance of RS Model Based on Graph Neural Network on MovieLens Dataset} 
\tiny
\begin{tabular}{cccccc}
\toprule[0.8pt]
\multirow{2}{*}{Dataset} & \multirow{2}{*}{year} & \multirow{2}{*}{Works}                                    & \multicolumn{2}{c}{Performance}                        & citations from                                  \\ \cline{4-6} 
                            &                       &                                                           & \multicolumn{1}{c}{NDCG}           & Recall            &                                                 \\ \midrule[0.8pt]
\multirow{2}{*}{1M}         & \multirow{2}{*}{2021} & \multirow{2}{*}{Xia et   al.\cite{xia2021incremental}}    & \multicolumn{1}{c}{NDCG@5 = 0.27}  & \multirow{2}{*}{} & \multirow{2}{*}{}                               \\ \cline{4-4}
                            &                       &                                                           & \multicolumn{1}{c}{NDCG@10 = 0.29} &                   &                                                 \\ \hline
\multirow{3}{*}{1M}         & \multirow{3}{*}{2022} & \multirow{3}{*}{Lin et al.   \cite{lin2022improving}}     & \multicolumn{1}{c}{NDCG@10 = 0.27} & Recall@10 = 0.21  & \multirow{3}{*}{}                               \\ \cline{4-5}
                            &                       &                                                           & \multicolumn{1}{c}{NDCG@20 = 0.28} & Recall@20 = 0.30  &                                                 \\ \cline{4-5}
                            &                       &                                                           & \multicolumn{1}{c}{NDCG@50 = 0.33} & Recall@50 = 0.47  &                                                 \\ \hline
\multirow{5}{*}{1M}         & \multirow{5}{*}{2020} & \multirow{5}{*}{He et   al. \cite{he2020lightgcn}}        & \multicolumn{1}{c}{NDCG@5 = 0.22}  & \multirow{2}{*}{} & \multirow{2}{*}{from~\cite{xia2021incremental}} \\ \cline{4-4}
                            &                       &                                                           & \multicolumn{1}{c}{NDCG@10 = 0.23} &                   &                                                 \\ \cline{4-6} 
                            &                       &                                                           & \multicolumn{1}{c}{NDCG@10 = 0.25} & Recall@10 = 0.19  & \multirow{3}{*}{from~\cite{lin2022improving}}   \\ \cline{4-5}
                            &                       &                                                           & \multicolumn{1}{c}{NDCG@20 = 0.26} & Recall@20 = 0.28  &                                                 \\ \cline{4-5}
                            &                       &                                                           & \multicolumn{1}{c}{NDCG@50 = 0.31} & Recall@50 = 0.45  &                                                 \\ \hline
\multirow{3}{*}{1M}         & \multirow{3}{*}{2019} & \multirow{3}{*}{Wang et   al.\cite{wang2019neural}}       & \multicolumn{1}{c}{NDCG@10 = 0.25} & Recall@10 = 0.18  & \multirow{3}{*}{from~\cite{lin2022improving}}   \\ \cline{4-5}
                            &                       &                                                           & \multicolumn{1}{c}{NDCG@20 = 0.26} & Recall@20 = 0.27  &                                                 \\ \cline{4-5}
                            &                       &                                                           & \multicolumn{1}{c}{NDCG@50 = 0.31} & Recall@50 = 0.43  &                                                 \\ \hline
\multirow{3}{*}{1M}         & \multirow{3}{*}{2019} & \multirow{3}{*}{Sun et   al.\cite{sun2019multi}}          & \multicolumn{1}{c}{NDCG@10 = 0.25} & Recall@10 = 0.18  & \multirow{3}{*}{from~\cite{lin2022improving}}   \\ \cline{4-5}
                            &                       &                                                           & \multicolumn{1}{c}{NDCG@20 = 0.26} & Recall@20 = 0.28  &                                                 \\ \cline{4-5}
                            &                       &                                                           & \multicolumn{1}{c}{NDCG@50 = 0.31} & Recall@50 = 0.44  &                                                 \\ \hline
\multirow{3}{*}{1M}         & \multirow{3}{*}{2020} & \multirow{3}{*}{Wang et   al.\cite{wang2020disentangled}} & \multicolumn{1}{c}{NDCG@10 = 0.25} & Recall@10 = 0.19  & \multirow{3}{*}{from~\cite{lin2022improving}}   \\ \cline{4-5}
                            &                       &                                                           & \multicolumn{1}{c}{NDCG@20 = 0.26} & Recall@20 = 0.28  &                                                 \\ \cline{4-5}
                            &                       &                                                           & \multicolumn{1}{c}{NDCG@50 = 0.31} & Recall@50 = 0.44  &                                                 \\ \hline
\multirow{3}{*}{1M}         & \multirow{3}{*}{2021} & \multirow{3}{*}{Wu et al.\cite{wu2021self}}               & \multicolumn{1}{c}{NDCG@10 = 0.25} & Recall@10 = 0.19  & \multirow{3}{*}{from~\cite{lin2022improving}}   \\ \cline{4-5}
                            &                       &                                                           & \multicolumn{1}{c}{NDCG@20 = 0.26} & Recall@20 = 0.28  &                                                 \\ \cline{4-5}
                            &                       &                                                           & \multicolumn{1}{c}{NDCG@50 = 0.31} & Recall@50 = 0.45  &                                                 \\ \bottomrule[0.8pt]
\end{tabular}
\end{table}

In the optimization of model-based CF, the selection of a suitable loss function plays a crucial role in determining the effectiveness of the system. Mean Square Error (MSE) and Log-likelihood are two commonly used methods to construct the loss function. When the rating data is explicit, MSE is typically applied, while when the rating is implicit, log-likelihood is used.
For instance, Li et al.~\cite{li2019capsule} used MSE as the loss function for their capsule neural network, comparing the predicted value to the true rating. Similarly, Seo et al.~\cite{seo2017interpretable} utilized MSE as the loss function for their recommender system, with a rating scale ranging from 1 to 5. On the other hand, Wu et al.~\cite{wu2019npa} treated the recommendation task as a pseudo $K+1$ way classification problem, using the log-likelihood function as their loss function with a rating scale of -1 and 1. It is important to note that the choice of loss function should align with the type of rating data available and the nature of the recommendation task at hand.

\subsubsection{Context-aware Based CF}
To enhance the decision-making capabilities of the recommendation system (RS), it is important to consider contextual information such as time, location, social status, mood, weather, day type, language, etc. This makes the RS more context-aware and leads to more accurate recommendations.
The relationship between user activity and contextual information is closely intertwined, as they both have a reciprocal effect on each other. This is why incorporating contextual information into the model results in a better recommendation.
There are various sources for obtaining contextual information. For instance, it can come from smart devices such as mobile phones that provide geolocation information (e.g., GPS) and companion data, or from the user directly through their personal interests~\cite{huang2022context}.

In recent years, advances have been made in the field of contextual information research. It can now be collected proactively by analyzing the user's activities on the system, instead of relying on self-reported information. This has been achieved through the application of statistical and data mining methods, especially artificial intelligent based methods~\cite{del2021ai}.

Generally, the contextual features are more complex and dynamic than user-item interaction data. For example, weather changes dynamically over time and it includes temperature, humidity, wind, precipitation, etc., with different scales.
Context information from smart devices is more complex, including, but not limited to, accelerometers, gyroscopes, magnetometers, GPS, biometric sensors, and so on, which can number up to 600 variables.
Therefore, considering all this contextual information in a recommender system is a challenging task.
To reduce its complexity, a direct approach is to use latent context embedding to reduce the dimension. Mei et al.~\cite{mei2018attentive} applied the interaction-centric module in their framework to obtain the interaction representation between contexts and users/items. Unger et al.~\cite{unger2017inferring} adopted the auto-encoder framework to learn contextual representations that might be useful for a recommendation. Recently, more contextual features are delved with complicated structures, Unger et al.~\cite{unger2019hierarchical} constructed a hierarchical model to obtain the latent representation of the context, which is the composition of a group similar latent context vectors, and it is also named contextual situation. 
Ouyang et al.~\cite{ouyang2022asymmetrical} believed that the existing symmetric context perception cannot fully extract features, so they designed an asymmetric context perception modulation and combined it with GNN to complete the recommendation.

We believe that the future contextual representation will be complex and dynamic with the new emergence of contextual information and complex models. 
Typically, there are three ways to include contextual information, based on the location and time of the deployment, it can be divided into pre-filtering, post-filtering, and contextual modeling~\cite{nawara2022context}. Pre-filtering is filtering out the irrelevant data before feeding context into the recommender system. Post-filtering is to apply the context to the recommender system after the standard recommendation (i.e., recommending without contextual information). Contextual modeling is to integrate the context into the recommendation algorithm directly. 
% Some works~\cite{panniello2009comparing} validate that post-filtering has a better performance than pre-filtering generally.
In a deep learning framework, using context information is rarely effective which leads to less work. To compensate for this gap, 
Ebesu et al.~\cite{ebesu2017neural} applied the convolutional neural network-based model to learn the contextual representation. 
% Unger et al.~\cite{unger2016towards} adopt an auto-encoder to extract latent contexts from a rich set of mobile sensors in an unsupervised manner. 
Jawarneh et al.~\cite{al2020pre} selected the contextual combinations by an item-splitting algorithm based on the t-statistics, and then apply the Prefiltering approach to incorporate the selected context into the neural collaborative filtering model.
Unger et al.~\cite{unger2020context} utilized the contextual modeling to incorporate the context, they first apply the auto-encoder to obtain the unstructured latent contextual representation from sensor data and then build a hierarchical tree from the learned representation to get the contextual situation which includes the time of the day, day of the week, location, and weather, etc., finally, the neural collaborative filtering model is applied to make a recommendation.
Wu et al.~\cite{wu2022graph} adopted the graph convolution machine which consists of the encoder, graph convolution layer, and the decoder to extract information from the user-item graph and its contextual information. 
To make a comprehensive review, all the relevant works that appear in recent years can be summarized into Table~\ref{tab:cars} based on the application, contextual information type, the way applied in context processing, and the model to make recommendations, etc.

\begin{table}
	\centering % centering table
		\label{tab:cars}
	\caption{Summarization of Contextual aware Recommender Systems.} % title name of the table
 \tiny
	\begin{tabular}{cccc}
		\toprule[0.8pt]
		 Application   & Contexts dimension  & Method         & Model Applied \\\midrule[0.8pt]
		     \makecell[c]{event and movie\\ recommendation~\cite{jhamb2018attentive} }      & \makecell[c]{some explicit \\contextual features} & \makecell[c]{Contextual \\modeling} & \makecell[c]{denoising \\autoencoder\\ neural network } \\\hline
		  \makecell[c]{food, mobile app \\recommendation~\cite{mei2018attentive}} & \makecell[c]{hungry status ( hungry, normal\\ and full), daytime, day \\of the week, and location} & Pre-filtering & \makecell[c]{attentive\\ interaction \\network}\\\hline
		   \makecell[c]{POI, movie, and \\music recommendation~\cite{unger2019hierarchical}}    & \makecell[c]{time, location, ringer mode,\\ battery, activity recognition,light,\\ accelerometer, orientation, etc.}     & \makecell[c]{Contextual \\modeling}  & \makecell[c]{matrix\\ factorization}   \\\hline
		         \makecell[c]{mobile app, music \\and movie recommendation~\cite{xin2019cfm} } & \makecell[c]{user ID, item ID, daytime,\\ the last music ID that\\ the user has listened \\within 90 minutes, movie ID \\and movie genres (multi-hot), etc.} &\makecell[c]{Contextual \\modeling}  & \makecell[c]{convolution \\neural network}    \\\hline
		  \makecell[c]{POI and movie\\ recommendation~\cite{al2020pre}}      & \makecell[c]{time, location, companion,\\ and travel type (family, couples,\\ business, solo travel, friends)}  & Pre-filtering    & \makecell[c]{neural \\collaborative \\filtering} \\\hline
		      \makecell[c]{POI and mobile app\\ recommendation~\cite{unger2020context}}     & \makecell[c]{location, noise level, \\part of the week, time of day, \\day of the week, etc. }  & \makecell[c]{Contextual \\modeling}  & \makecell[c]{neural \\matrix\\ factorization} \\\hline
		  e-commerce~\cite{wu2022graph} & \makecell[c]{time, time difference, \\ event type (view, sale or\\add to basket), etc.} & \makecell[c]{Contextual \\modeling}  & \makecell[c]{graph convolution,\\ encoder, decoder}\\\bottomrule[0.8pt]
	\end{tabular}
\end{table}

\subsection{Content Based RS}
\label{subsec:contentrs}
Unlike collaborative filtering, content-based RS treats recommendations as a user-specific classification problem by learning a classifier based on item features, and it provides a more personal recommendation. The system mainly focuses on two types of information which are:
\begin{itemize}
	\item user preference model; and
	\item user interaction history. 
\end{itemize}

With this information, the RS obtains the feature of the items with the item presentation algorithm. Text data is one of the important sources of feature learning and is the implicit data in the recommender system that is beneficial to both the user and the item. 
One advantage is that we can only make recommendations based on the content contained in the item, without rating. 
The side information in the content-based RS can be metadata or user-generated content. Traditionally, it is the descriptions of items or full-text indexing of textual items. However, a great deal of work has been done to extend metadata with external knowledge to make it knowledge-based RS, as described in the section~\ref{sec:kbs}.
In addition to textual modality, visual and multimedia features are becoming more popular in content-based RS~\cite{deldjoo2020recommender}. 
Thus, deep learning especially the techniques in natural language processing can be applied in the content-based RS.
For example, Van et al.~\cite{van2022solving} concatenated the word embedding with other personal favors to learn user profiles, and experimental results show that this method was more effective than collaborative filtering.
Yang et al.~\cite{yang2022personalized} extracted additional features and side information from an open knowledge graph with an autoencoder to make better recommendations. Deldjoo et al.~\cite{deldjoo2018using} applied a pre-trained neural network to obtain the stylistic properties from text, which is helpful for a better recommendation.
Cami et al.~\cite{cami2019user} applied the posterior inference to create the user model based on the user profile and make a better recommendation with the latent clustering method.
Wang et al.~\cite{wang2019camo} applied GRU to construct the content encoder and build the model CAMO in the text feature extraction. With the help of adversarial training, it has gained state-of-the-art results in the recommendation. Jesus et. al.~\cite{aspect_jesus2020} built a method for determining personal preferences and a variation coefficient approach for assigning significance to the product review aspects. This research presents an adaptable system for incorporating essential features in the design of a customized recommender system, considering both the opinion providers and the ultimate user. Marco et. al.~\cite{POLIGNANO2021114382} contributed by introducing a comprehensive emotion-sensitive computational model, which is based on the creation of affecting user profiles. Each preference, such as a 5-star rating for a movie, is linked to the emotional state experienced by the user at the time the preference was recorded. The emotional state of the user, comprising a range of emotions such as joy and surprise, is a crucial aspect of the decision-making process and must be considered in modeling user preferences.
Ray et al.~\cite{RAY2021106935} presented a hotel recommendation system that leverages sentiment analysis of hotel reviews and aspect-based review categorization, based on user queries. Additionally, they introduced a new extensive and diverse dataset of online hotel reviews collected from Tripadvisor.com.

In content-based RS, the Heterogeneous Information Network (HIN) approach is widely used to connect different types of side information in a parallel manner, as many recommender systems involve multiple item types and related links.
HIN is a unique network that incorporates multiple item types and relationships to effectively integrate data from one or more domains, adding richer semantics to items and links. These networks can take the form of social networks, knowledge graphs, or other types of information networks. For example, 
Amato et al.~\cite{amato2019sos} created a user-content social graph based on multimedia content, using semantic relevance and low-level feature comparisons to make recommendations through ranking technology.
Ma et al.~\cite{ma2019newly} built a heterogeneous bibliographic network using citation information and made recommendations based on meta-paths in the HIN.
Pham et al.~\cite{pham2023hierarchical} proposed the fuzzy-driven HIN embedding model for the recommendation which integrated hierarchical structure from fuzzy and deep network-based embedding architecture.
Later, they applied the HIN to extract features from unstructured attributes or contents~\cite{pham2023approach}.
If the HIN is a knowledge graph, it will be further explained in Section~\ref{sec:kbs}.

In the recommendation process, either clustering or ranking-based methods can be used. Typically, clustering is based on the Bayesian model, while the ranking is based on the distance between candidates' and users' preferences. Different similarity metrics, including Kendall-Tau, Spearman, Pearson, etc., can be used for similarity calculation. These metrics will be thoroughly discussed in Section~\ref{subsec:sim}.

\subsection{Knowledge-based RS}
\label{sec:kbs}
\begin{figure}[ht]
	\centering
	\includegraphics[width=0.9\linewidth]{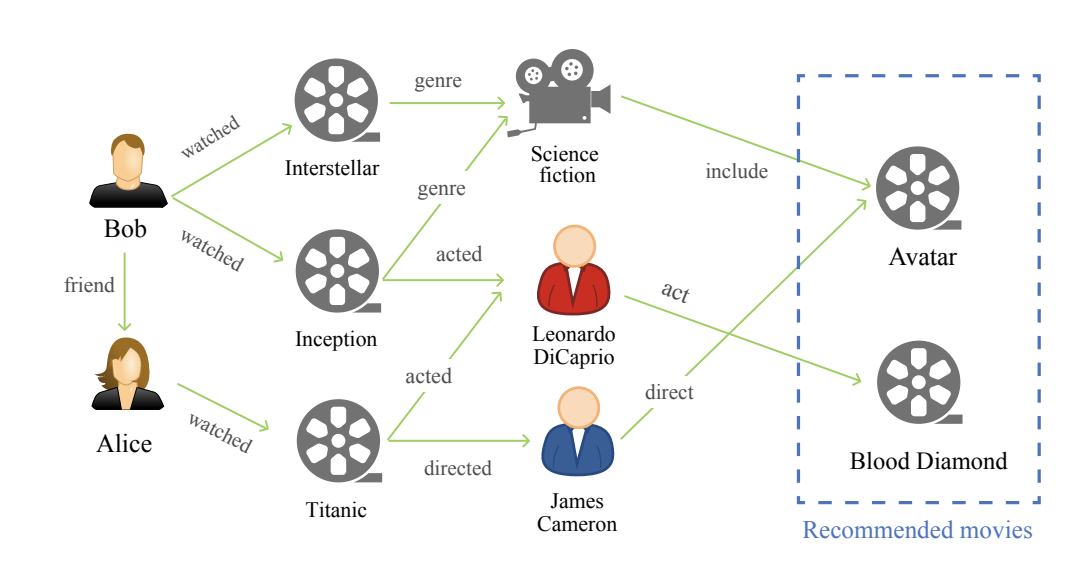} 
	\caption{An example of a knowledge-based recommender system.(Source:~\cite{chen2019towards}.)}
	\label{fig:example_kbrs}
\end{figure}

The knowledge-based RS operates differently from CF-based and content-based RS as it does not rely on user-item interaction history. It is commonly used in complex domains where items are not typically purchased, such as cars and department businesses. In these scenarios, users may have constraints, such as a budget limit, when making a purchase. The key information for knowledge-based RS  is domain knowledge, which can encompass user knowledge, item knowledge, or the interaction between users and items~\cite{chen2019towards}. This information helps to overcome challenges such as cold start and grey sheep~\cite{alabdulrahman2021catering}. By incorporating domain knowledge, knowledge-based RS  can overcome the challenges posed by grey sheep and provide more accurate recommendations.
As shown in Fig.~\ref{fig:example_kbrs}, domain knowledge can be located on both the user and item sides.

The advancement of Knowledge-based RS has garnered significant attention in recent years due to its versatility in a wide range of domains. By integrating contextual knowledge of items into multi-turn dialog systems, Knowledge-based RS has been able to improve the natural language generation process as demonstrated by Chen et al.~\cite{chen2022multimodal}. In the e-commerce sector, Vijayakumar et al.~\cite{vijayakumar2019effective} utilized user-selected Point-of-Interest information as temporal features to personalize travel plans. In the field of news recommendation, Liu et al.~\cite{liu2021reinforced} enhanced news recommendation reasoning based on a knowledge graph combined with reinforcement learning. Huang et al.~\cite{huang2021knowledge} proposed the Knowledge-aware Coupled Graph Neural Network (KCGN), which incorporates inter-dependent knowledge across items and users into the recommendation framework. This enhances the encoding of high-order user- and item-wise relationships, as well as the utilization of mutual information for global graph structure awareness. Wang et al.~\cite{wang2021learning} explored the intents behind user-item interactions using auxiliary item knowledge, proposing the Knowledge Graph-based Intent Network (KGIN). This model models each intent as an attentive combination of KG relations, promoting independence among different intents for better model capability and interpretability. 
Xia et al.~\cite{xia2021knowledge} introduced the Knowledge-Enhanced Hierarchical Graph Transformer Network (KHGT) to explore multi-typed interactive patterns between users and items in recommender systems. The model is based on a graph-structured neural architecture, which focuses on capturing type-specific behavior semantics, as well as explicitly discriminating the importance of different types of user-item interactions.

In the medical sector, Gong et al.~\cite{gong2021smr} proposed a novel framework, Safe Medicine Recommendation (SMR), that builds upon recent advances in graph embedding learning techniques. SMR first creates a high-quality heterogeneous graph by connecting Electronic Medical Records (EMRs) and medical knowledge graphs, and then jointly embeds diseases, medicines, patients, and their corresponding relationships into a shared lower-dimensional space. In software development, Bilal et al.~\cite{abu2021toward} utilized query expansion to enhance the recommendation of personnel development tools during website searches.
Cui et al.~\cite{cui2022ketch} enhanced the health thread recommendation method based on a knowledge graph, focusing not only on the user's interaction with the thread but also on the underlying conditions and symptoms.

Overall, these studies demonstrate the versatility and potential of knowledge-based RS in different domains, such as e-commerce, healthcare, academia, and software development. These applications demonstrate the potential of integrating domain knowledge into RS to make more informed and personalized recommendations.

\subsubsection{Knowledge Graph-based RS}
Knowledge graph-based RS is a sub-class of knowledge-based RS that leverages the information contained in a knowledge graph to make better recommendations. The knowledge graph is a collection of entities and relationships between them, including objects, relations, and situations. It provides additional information to the recommender system and enhances the quality of recommendations. The application of knowledge graphs in RS has gained significant attention in recent years, becoming a popular research topic. By integrating the knowledge graph into the recommender system, more evidence can be gathered to improve the quality of recommendations.

There are several existing knowledge graphs, such as YAGO~\cite{suchanek2007yago}, Freebase~\cite{bollacker2008freebase}, DBpedia~\cite{lehmann2015dbpedia}, NELL~\cite{carlson2010coupled}, Wikidata~\cite{vrandevcic2014wikidata}, etc., which cover a wide range of domains. Additionally, there are some specialized knowledge graphs, such as Bio2RDF~\cite{belleau2008bio2rdf} for biology, Knowlife~\cite{ernst2014knowlife} for biomedical information, and Satori~\cite{gao2018building} for production. Table~\ref{tab:kg_detail} provides further details on these knowledge graphs.

\begin{table}[htp]
	\centering
		\caption{The details of the existing knowledge graph.}
	\label{tab:kg_detail}
 \tiny
	\begin{tabular}{cc}
		\toprule[0.8pt]
		Knowledge graph & Domain      \\\midrule[0.8pt]
		YAGO~\cite{suchanek2007yago}          & Cross Domain \\
		DBpedia~\cite{lehmann2015dbpedia}        & Cross Domain \\
		Freebase~\cite{bollacker2008freebase}       & Cross Domain  \\
		NELL~\cite{carlson2010coupled}           & Cross Domain  \\
		Wikidata~\cite{vrandevcic2014wikidata}  & Cross Domain  \\
		Satori~\cite{gao2018building}             & Production  \\
		Bio2RDF~\cite{belleau2008bio2rdf}       & Biological  \\
		Knowlife~\cite{ernst2014knowlife}       & Biomedical  \\\bottomrule[0.8pt]
	\end{tabular}

\end{table}

A knowledge graph constitutes an entity-relation-entity triples $(h,r,t)$, where $h$ is the head, $r$ is the relation, and $t$ is the tail. 
There are three different ways in the knowledge graph utilization~\cite{guo2020survey,balloccu2022post} in a recommender system, namely knowledge graph embedding, path pattern learning, and hybrid embedding methods.

\textbf{Knowledge Graph Embedding:}
First, it can be used as the embedding by encoding existing knowledge into a higher-level representation which is named knowledge graph embedding (KGE). Then it is aggregated as the side information in the latent vector $v_{j}$ for item $j$. Most of the works are about the transition-based KGE, and such equation $\mathbf{h}+\mathbf{r}\approx \mathbf{t}$ will be followed if a triple exists in the knowledge graph, where $\mathbf{h},\mathbf{r},\mathbf{t}$ are the vectors for the head, relation, and tail.
Based on different mapping methods, the existing transition-based KGE score function is summarised as follows:
\begin{itemize}
	\item TransE~\cite{bordes2013translating} learns the embedding directly by forming score function $
	f_{s}(h,t) = \|\mathbf{h}+\mathbf{r}-\mathbf{t}\|^{2}_{2}
	$. If $f_{s}$ is low, the relation in $(h,r,t)$ will hold, otherwise will not.
	\item TransH~\cite{wang2014knowledge} learns the embedding in the relation hyper-planes by mapping the entity with $\mathbf{h_{H} = \mathbf{h} - \mathbf{w}^{T}_{H}\mathbf{h}\mathbf{w}_{H}}$, $\mathbf{t_{H}} = \mathbf{t} - \mathbf{w^{T}_{H}}\mathbf{t}\mathbf{w_{H}}$ where $\|\mathbf{w_{H}}\|_{2}=1$. Then the score function is calculated as $f_{s}(h,t) = \|\mathbf{h_{H}}+\mathbf{r} - \mathbf{t_{H}}\|^{2}_{2}$.
	\item TransR~\cite{lin2015learning} maps the entity embedding with projection matrix $\mathbf{M}_{r}$ for the relation $r$, then we have $\mathbf{h_{r}} =\mathbf{h} \mathbf{M}_{r}$ and $\mathbf{t_{r}} =\mathbf{t}\mathbf{M}_{r}$. The score function is calculated as $f_{s}(h,t) = \|\mathbf{h_{r}}+\mathbf{r}-\mathbf{t_{r}}\|_{2}^{2}$.
	\item TransD~\cite{ji2015knowledge} uses two projection vectors to transfer the entity-relation pair: $\mathbf{h_{d}} = (\mathbf{r_{p}}\mathbf{h_{p}^{T}} +\mathbf{I})\mathbf{h}$, $\mathbf{t_{d}} = (\mathbf{r_{p}}\mathbf{t_{p}^{T}} +\mathbf{I})\mathbf{t}$, in the transfer equations, $\mathbf{h_{p}},\mathbf{r_{p}},\mathbf{t_{p}}$ are another set of vectors for the entities and relations, $\mathbf{I}$ is the identity matrix. The score function is calculated as $f_{s}(h,t) = \|\mathbf{h_{d}} + \mathbf{r} - \mathbf{t_{d}}\|_{2}^{2}$.
	\item DistMult~\cite{yang2014embedding} applies the bi-linear equation in the score function construction, and it can be written as $f_{s}(h,t) = \mathbf{h}\mathbf{M}_{r}\mathbf{t}$, where $\mathbf{M}_{r}\in R^{n\times n}$.
\end{itemize}
Based on the score function, the margin-based ranking loss is always applied in the objective function which is written in Equation~\ref{equ:loss_margin}.
\begin{equation}
	\label{equ:loss_margin}
\small
	L = \sum_{(h,r,t)}\sum_{(h',r',t')} \max(0, f_{s}(h,t) +\gamma- f_{s}(h',t'))
\end{equation}

where $(h,r,t)$ is the correct knowledge triple, while $(h',r',t')$ is the craft incorrect knowledge triple, and $\gamma$ denotes the margin between the triples.

The field of external knowledge addition in KGE (knowledge graph embedding) has seen a number of recent advancements. Zhang et al.~\cite{zhang2016collaborative} incorporated structural, visual, and textual knowledge into their model by transferring the knowledge into embeddings. Wang et al.~\cite{wang2018dkn} proposed a deep knowledge-aware network for news recommendation, using the TransD~\cite{ji2015knowledge} algorithm to transfer knowledge embeddings from the Satori knowledge base.
Huang et al.~\cite{huang2018improving} improved sequential recommendation by incorporating external knowledge information into their key-value memory network. They used the TransE~\cite{bordes2013translating} algorithm to transfer the knowledge embedding, using Freebase~\cite{bollacker2008freebase} as the source of external knowledge. Similarly, Dadoun et al.~\cite{dadoun2019location} utilized the TransE~\cite{bordes2013translating} algorithm to transfer the knowledge graph from Wikidata~\cite{vrandevcic2014wikidata} into their deep knowledge factorization machines for POI recommendation.
Recently, Hu et al.~\cite{hu2022transmkr} made further progress in this area by using the TransR~\cite{lin2015learning} algorithm to quantify relationships between POIs and their attributes. This method allowed for the cross-sharing of information between POI and entity vectors, thus improving the expressive ability of POI data and addressing the issue of data sparsity. Huang et al.~\cite{huang2021entity} designed an Entity-Aware Collaborative Relational Network (ECRN) to extract interactive entity semantic relations. Tu et al.~\cite{tu2021conditional} designed the knowledge-aware conditional attention network (KCAN) to refine the useful information of the knowledge graph to better improve the performance of RS. 
Anelli et al.~\cite{anelli2021sparse} designed the KGFlex model based on the knowledge map to obtain sparse features of interaction and context and made recommendations on this basis. Yang et al.~\cite{yang2022knowledge} designed a contrastive learning model using knowledge graphs, which can learn perceptual representations with less noise for the project. Experiments have proved that the model has a better anti-noise ability and ability to deal with long-tail effects. Wang et al.~\cite{wang2022multi} designed a multi-level reasoning framework based on knowledge graphs to learn users' multi-level interests, and explained the reasoning path through path selection. 
Geng et al.~\cite{geng2022path} proposed the Path Language Modeling Recommendation (PLM-Rec) framework by learning a language model to solve the problem of knowledge-based RS recall bias. 
Park et al.~\cite{park2022reinforcement} combined user sentiment analysis on the basis of knowledge graphs to improve the performance and interpretation quality of RS.
Overall, these works demonstrate the use of aggregation methods in the addition of external knowledge in KGE.

\subsubsection{Path-based RS}
The path-based method in recommendation systems leverages the connectivity patterns between entities in a knowledge graph to enrich the side information in the users or items. This is done by evaluating the semantic similarity between users or items through different meta-paths. There are three types of entity similarity calculation including User-User, Item-Item, and User-Item similarity~\cite{qinsurvey}.

In~\cite{ma2019jointly}, Ma et al. leveraged an external knowledge graph to learn association rules and integrate these rules into the interaction matrix through matrix factorization for recommendation generation. On the other hand, Xu et al.~\cite{xu2021recommendation} adopted a TransE-based approach to embed target users and items and combine knowledge graph and reinforcement learning to determine the appropriate recommendation items and discover reasoning paths from target users to the recommended items. Moreover, in~\cite{wu2022knowledge}, the authors proposed a knowledge graph-based multi-context-aware recommendation algorithm that blends the benefits of both path-based and propagation-based methods. They also introduced a novel concept, referred to as ``rule'', to characterize the user's preferences.

Sun et al.~\cite{sun2018recurrent} employed a recurrent network in their model, Recurrent Knowledge Graph Embedding (RKGE), to encode the path linking the same entity pair and learned the semantic representations of both entities and paths to characterize user preferences towards items. Wang et al.~\cite{wang2019explainable} proposed the Knowledge-Aware Path Recurrent Network (KPRN) that used LSTM to encode the path sequence with relation and entity embeddings to estimate the user's preference for a recommendation.
Zheng et al.~\cite{zheng2022explainable} proposed a guided model based on meta-path and self-attention mechanism, which can provide a reasonable explanation for session-based recommendation (SR). Zhao et al.~\cite{zhao2022two} designed a multi-turn dialogue recommendation model GPR based on graph paths, and made recommendations through the optimal paths learned by soft clustering, set operations, and reinforcement learning. Ning et al.~\cite{ning2022automatic} designed a meta-path selection framework (RMS) for heterogeneous graphs based on reinforcement learning and proposed a new meta-path recommendation for its RMS HRec, which is superior to existing models.

In conclusion, the path-based method has been widely applied in recommendation systems to leverage the connectivity patterns in the knowledge graph for better recommendation results. Different models have used different similarity evaluation methods, such as PathSim, Hetesim, and LSTM, to calculate the semantic similarity between entities and generate recommendations.

\subsection{Hybrid RS}
\label{subsec:hybrs}
The development of recommender systems has resulted in increased complexity, particularly in their practical applications. To achieve real-time, accurate recommendations, these systems often require the integration of various data sources. The use of deep learning has allowed for the more effective fusion of multi-source information, such as context and content, in hybrid recommender systems~\cite{kiran2020dnnrec}. Polignano et al.~\cite{polignano2021together} blend user-item interactions and contextual features for the recommendation. Luo et al.~\cite{luo2022hysage} mixed graph embedding and context information for the recommendation.

To provide a comprehensive overview of the hybrid recommender system, a table summarizing related works is presented in this section. Table~\ref{tab:hybrid_summary} categorizes these works based on their ``core model'', ``composition'', ``information source'', and ``applications''. This table provides a useful reference for researchers and practitioners interested in the field of hybrid recommender systems.

\begin{table}[htp]
	\tiny
	\caption{The summary about the hybrid recommender system based on the model and data source.}
	\label{tab:hybrid_summary}

	\begin{tabular}{cccc}
		\toprule[0.8pt]
		 Core Model                         & Composition                  & Information Source                                                 & Application           \\\midrule[0.8pt]
		                \makecell[c]{Graph Model \\(LCP theory)~\cite{aslan2019hybrid} }                 & \makecell[c]{Content based, \\ CF based}           & \makecell[c]{Content (Papers),  \\Context (Authors' \\social network)}                                & \makecell[c]{Researcher \\Recommendation}    \\\hline
		            Graph Model~\cite{gatzioura2019hybrid}                         & \makecell[c]{Content-based,\\ CF based}         &\makecell[c]{ Content (Music\\ Descriptions),  Context\\ (User Preferences) }                             & \makecell[c]{Music \\Recommendation}       \\\hline
		            \makecell[c]{Hybrid Probabilistic \\Graphical Model~\cite{kouki2019personalized} }          & \makecell[c]{Content-based, \\CF based}           & \makecell[c]{Content (Music Style),  \\Context(User's history,\\ social connections, tags, \\and popularity statistics) }       & \makecell[c]{Music \\Recommendation }      \\\hline
		\makecell[c]{}   AutoEncoder~\cite{zhu2022variational} ~\cite{strub2016hybrid} ~\cite{liu2018novel}                 & CF based                    &\makecell[c]{ Content (User, Item), \\Context  (User's ID, age, gender, \\occupation, zip code, \\movie genres, title) }         & \makecell[c]{General \\Recommendation}      \\\hline
		                DNN~\cite{kiran2020dnnrec}                            & CF based                    & \makecell[c]{Content (User, Item), \\ Context(Movie genres,\\ title, user tags) }                          & \makecell[c]{General \\Recommendation}      \\\hline
		% Ma et al.~\cite{ma2018partial}                & VAE                            & \makecell[c]{Content-based, \\CF based }            & Content (User, Item)                                                & General Recommendation      \\\hline
		                SVD ~\cite{walek2020hybrid}                           & \makecell[c]{Content-based, \\CF based, Fuzzy \\ Expert System} & \makecell[c]{Content (User, Movie),  \\Context(Average movie \\rating, number of ratings)}                      & \makecell[c]{Movie \\Recommendation}       \\\hline
		% Wang et al.~\cite{wang2018adapting}              & \makecell[c]{Breadth-First Search \\over Social  Influence \\Network}    & Content-based                 & \makecell[c]{Content (User, Business), \\ Context (Social network) }                                & Business Recommendation     \\\hline
		              \makecell[c]{Alternating Least Squares, \\ Bayesian Personalized \\Ranking ~\cite{barros2020hybrid}}& \makecell[c]{Content-based, \\CF based }           & \makecell[c]{Content (Authors from \\research  articles, \\Chemical Compounds \\in ChEBI,  Number \\of articles the author \\ wrote)} & \makecell[c]{Chemical Compound\\ Recommendation} \\\hline
		              \makecell[c]{Alternating Least Squares, \\Collaborative Filtering, DNN~\cite{biswas2022hybrid} }& \makecell[c]{Content-based, \\CF based }           & \makecell[c]{Content (User Preferences, \\Browsing/Search/clicks/\\order history} & \makecell[c]{Business\\ Recommendation} \\\bottomrule[0.8pt]
	\end{tabular}
\end{table}

The table shows that the majority of works utilize a CF-based model as the optimization framework. In addition to content information, contextual information includes a variety of features such as historical behavior, social connections, and personal characteristics. These findings highlight the importance of incorporating multiple data sources in the development of effective recommender systems.
 
\section{Group Recommender Systems}
The field of recommendation systems has seen a significant amount of research in recent years, with the majority of work focusing on individual recommendation systems. However, there is a growing body of literature that deals with group recommendation systems, which aim to provide recommendations to a group of users who have similar preferences, as opposed to individualized recommendations.

\begin{figure}[ht]
	\centering
	\includegraphics[width=0.8\linewidth]{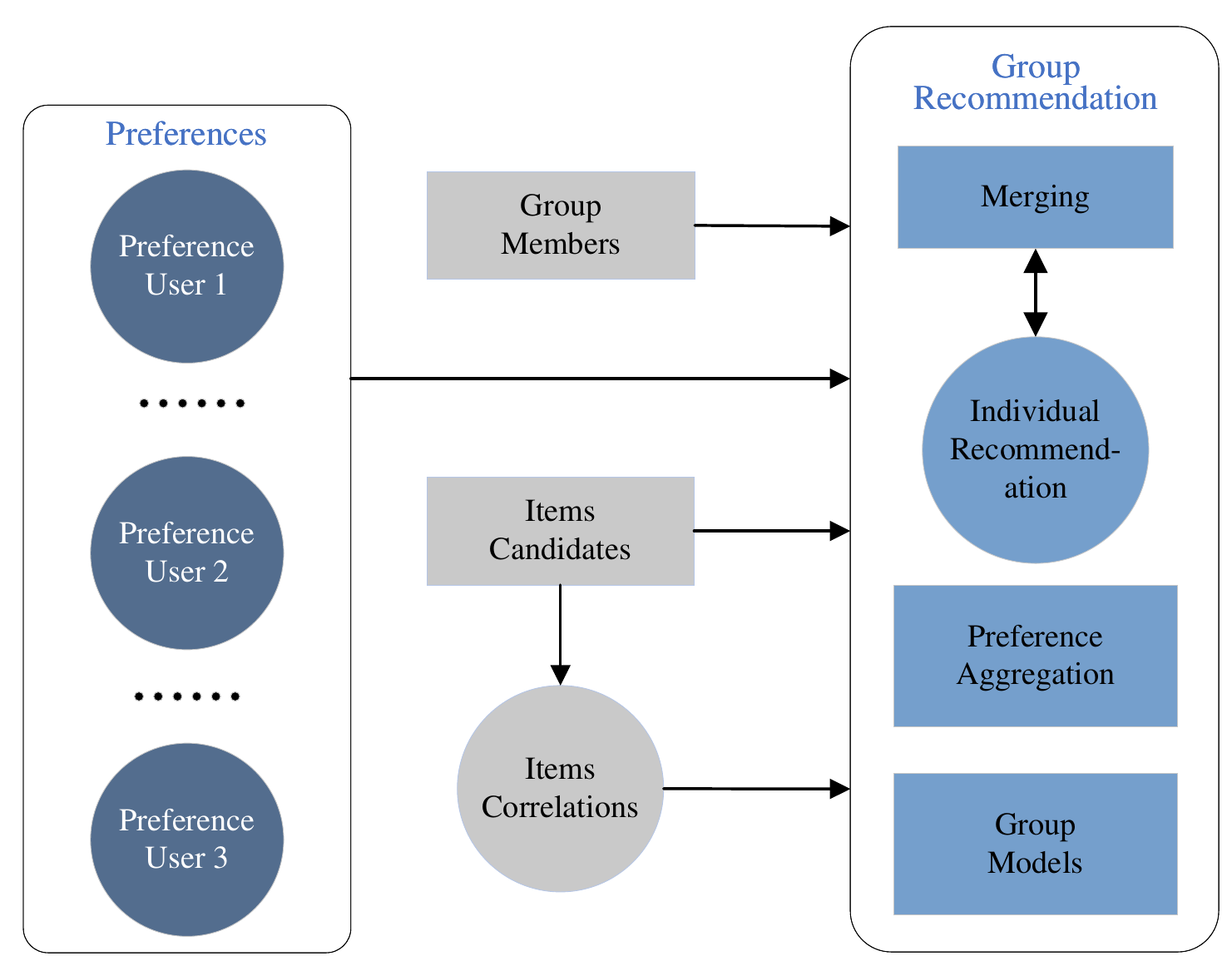}
	\caption{A general framework of group recommender system. (The figure is from~\cite{dara2019survey}.)}
	\label{fig:example_grs}
\end{figure}

A general framework for group recommendation systems is shown in Fig.~\ref{fig:example_grs}. This framework highlights the key components involved in generating a group recommendation, such as group creation and recommendation generation.
To create a group, various factors can be considered, including shared interests, activities performed together, or being in the same environment. The group creation process can also be influenced by the aggregation of recommendations, as described in~\cite{stratigi2022sequential}.

In order to generate a good group recommendation, it is important to first create the group and then generate the recommendations, as described in~\cite{huang2021novel}. This involves considering the factors mentioned above and using them to create a group of users with similar preferences.

It is worth noting that the field of group recommendation systems is still in its early stages of development, and further research is needed to fully understand its potential and limitations. Despite this, the growing body of literature in this area highlights its importance, and it is likely that group recommendation systems will become increasingly common in the near future.

Based on the number of interactions between group members, it can be divided into the memory-based approach and the model-based approach. 
There is no or minimal interaction among the group in the memory-based approach, and they aggregate the preference only. 
While the model-based approach tries to learn a comprehensive representation of the group with machine learning methods. And the recommended performance of the model-based is better than the memory-based approach in general.

\subsection{Memory-based Approach}
The most commonly used method is preference aggregation in a group recommendation. 
Some works divide preference aggregation into result aggregation and profile aggregation by the way they make group recommendations~\cite{wu2022graph}.
Result aggregation is to aggregate all users' recommendation results together to make a comprehensive recommendation for the whole group. 
Recently, with the development of the graph neural network, group recommendation 
For example, Pitoura et al.~\cite{pitoura2022fairness} adopted the rank aggregation methods in group recommendation after giving recommendations to the individual users.
While profile aggregation is to make personal recommendations to a virtual user who is abstracted from the profile of group members. For example, 
Wang et al.~\cite{wang2019trugrc} proposed a trust-aware group recommendation model by combing the result aggregation and the profile aggregation together.
Generally, most of the works make group recommendations after they finished the aggregation of the preferences.

There are different aggregation strategies, in the work~\cite{huang2021novel}, and they classified these methods into majority-based strategies (e.g. plurality voting), consensus-based strategies (e.g. average~\cite{pitoura2022fairness}, average without misery and fairness~\cite{malecek2021fairness}), and borderline strategies (e.g. least misery~\cite{kumar2022automatically}, and most pleasure~\cite{zan2021uda}).
In the majority-based strategies, plurality voting applies the rule of first past the post and chooses the most votes item repetitively, for example, it will choose the highest votes items first, then the second highest votes item. 
In consensus-based strategies, 
Averages mean the item is selected according to the average rating of the group;
Average without misery means the rating that is below a certain threshold will be excluded from the average individual ratings;
Fairness means items will be ranked in a way that individuals are choosing them in turn, the people who choose first will have the most satisfaction.
In the borderline strategies, Least misery refers to the minimum individual rating in a group, assuming that the least satisfied member represents the group's preference;
While, most pleasure is to capture the maximum rating in a group, assuming that the most satisfied member represents the group's preference.
Each of these strategies has pros and cons, for example, the Average strategy will perform well when the group's preferences are quite similar, but it will lead to worse recommendations when the preferences are diverse. In the work~\cite{huang2021novel}, they provided a detailed description of the strategies selection. And Sato et al.~\cite{sato2022enumerating} proposed a method to support filtering queries to effectively enumerate fair packages to alleviate the fairness problem in group recommendation.

\subsection{Model-based Approach}
Unlike memory-based approaches, model-based approaches rely on machine learning techniques such as information fusion, game theory, and probabilistic models to extract relevant features from user interactions. The deep learning method, in particular, has become a crucial component in group recommendation as it allows for the integration of various external features, such as social features, group descriptive features, and commonsense knowledge, which can enhance recommendation performance.

Several studies have explored the use of deep learning in group recommendation. For instance, Yin et al.~\cite{yin2019social} aggregated personal preferences and social influence to generate recommendations for occasional groups. Zhu et al.~\cite{zhu2018group} generated group recommendations using a knowledge map and clustering method. Cao et al.~\cite{cao2018attentive} used an attention network and neural collaborative filtering to learn the aggregation strategy directly from the data. The attention mechanism, which has a natural advantage in feature aggregation, has become a popular technique in group recommendation. Huang et al.~\cite{huang2020efficient} applied the multi-attention mechanism to the deep-learning method in group recommendation, while Wang et al.~\cite{wang2020group} combined sequential and group recommendations using a graph representation approach. In~\cite{wang2019group}, the self-attention mechanism is used to learn the weight of preferences between group members, which is then aggregated to generate recommendations. Aiming at the severely sparse group-item interaction in group recommendation, Zhang et al.~\cite{zhang2021double} proposed a self-supervised hypergraph learning framework, which includes a hierarchical hypergraph convolutional network based on user-level and group-level hypergraphs and a dual-scale node loss strategy. Create self-supervised signals, the former used to capture intra-group and inter-group interactions, and the latter to alleviate the sparsity of the original data. Sajjadi et al.~\cite{sajjadi2021deepgroup} defined the cold start problem of group recommendation as group recommendation from group implicit feedback, considered the user's personal privacy issues and proposed a DeepGroup model that uses group implicit data for group recommendation. Zhang et al.~\cite{zhang2022gbert} proposed the GBERT pre-training and fine-tuning method for the temporary group recommendation problem. The expression ability of BERT is enhanced to capture group preferences. In the pre-training stage, three tasks are used to alleviate the sparsity and cold-start problems, and the adjustment target fine-tuning weight is designed. Chen et al.~\cite{chen2022thinking} believed that there is a problem with representing user and group preferences with points in space, and proposed a CubeRec model based on hypercubes, by learning group hypercubes from user embedding during preference aggregation, with improved distance metric performance, group hypercubes and affinity between item points, addressing data sparsity by crossing two groups combined with self-supervision.

Despite the numerous deep learning-based methods proposed for group recommendation, the recommendation performance still remains unsatisfactory due to the sparse and diverse preferences of group users. Further research is needed to address these challenges and enhance the recommendation performance.
\subsection{Bundle Recommendation}
Bundled recommendations can reduce the time spent by users and increase sales revenue while meeting user needs by selling two or more items as a whole. The most typical example is the ``beer diaper case'' of Wal-Mart in the 1980s, that is, putting diapers and beer in the same place at a specific time, which not only increased operating income but also saved users' shopping time. Agrawal et al.~\cite{agrawal1993mining} proposed the mining rule, which can be regarded as the theoretical basis for bundling recommendations.

The purpose of group recommendation is to recommend items to a group of people who have something in common, while the purpose of the bundled recommendation is to recommend a bundled item to users as a whole. Therefore, group recommendation and bundled recommendation have a lot in common. We can think of bundled recommendations as recommendations for groups of items.

Bai et al.~\cite{bai2019personalized} proposed a beam generation network (BGN), which decomposes the structured prediction problem into quality and diversity parts with a determinant point process (DPP) and alleviates insufficient expression through the encoder-decoder framework and the proposed feature-aware softmax, through Masked beam search and DPP selection for the recommendation. Based on the graph convolutional neural network, Liu et al.~\cite {liu2020basconv} proposed a new framework of BasConv. Through the integration of three aggregators that can distinguish different types of interactions in the framework, the aggregator can learn node embedding expressions from neighborhoods and high-order neighborhoods. He et al.~\cite{he2022bundle} proposed the BundleMCR recommendation task for multi-round conversation recommendation (MCR), and proposed the BundleBert (Bunt) model that uses the Markov decision process of multiple agents to recommend. Zhao et al.~\cite{zhao2022multi} proposed the Multi-View Intent Separation Graph Network (MIDGN) model, which separates user intent from two perspectives, global view and local view, and accurately and comprehensively captures the diversity of user intent and item associations at a finer granularity. Avny et al.~\cite{avny2022bruce} proposed the BRUCE model based on Transformers to capture the relationship between an item and other items in the bundle, improving the accuracy of the recommendation.

\section{Problems in Recommender System}
Despite its popularity and success, RS still faces several challenges that need to be addressed in order to improve its performance and usability. Three major challenges faced by RS are robustness, data bias, and fairness.
Robustness refers to the ability of RS to withstand adversarial attacks and maintain its performance under such conditions, and the ability to eliminate the adverse effects of bias on RS data or models. Adversarial attacks can take different forms, such as manipulating data, model parameters, or even the output of RS. Therefore, ensuring the robustness of RS is crucial to guarantee its safety and reliability.
Data bias means that the training data only provides a skewed snapshot of user preferences, making the recommendation model fall into suboptimal results~\cite{chen2020bias}. The bias may include popularity bias, selection bias, exposure bias, position bias, and other biases. 
Fairness means that the recommendation results of RS are based on the principle of impartiality and fairness, which includes user-based fairness and project-based fairness. User-based fairness means that the recommended results are not affected by sensitive attributes (such as race, gender, etc.) that may lead to discrimination. Item-based fairness means that the recommendation process should try to ensure that similar items of the same category have equal chances to be recommended. Typical cases of fairness problems are cold start problems and long tail problems.   

In this survey paper, we aim to provide a comprehensive overview of the challenges faced by RS in terms of robustness and fairness. We discuss various existing methods and techniques that address these challenges and evaluate their strengths and limitations. Through this paper, we aim to provide insights into the current state of the field and suggest future directions for research in this area.

\subsection{Robustness}

Adversarial attacks on recommender systems have become a popular method for validating the robustness of these systems. Researchers often add perturbations to the pixel values or text in image processing or natural language processing systems, respectively. In natural language processing, an item or user can be represented as an embedding, which provides attackers with more options to attack, including utilizing information such as user profile, item profile, user embedding, item embedding, and side information.

Adversarial attacks can be further categorized into black-box attacks and white-box attacks. In a white-box attack, the attacker has access to the model information and uses gradient-based methods such as FGSM, PGD, and C\&W to search for adversarial perturbations. In a black-box attack, the attacker does not have access to the model information and uses methods such as model substitution or DeepFool to attack. For example, He et al.~\cite{he2018adversarial} added perturbations to the user and item embeddings using FGSM and were the first to perform an adversarial attack on a recommender system. Similarly, Tang et al.~\cite{tang2019adversarial} added perturbations to the side information using FGSM, and Christakopoulou et al.~\cite{christakopoulou2019adversarial} added perturbations to the user profile using a generative adversarial network-based method. Zhang et al.~\cite{zhang2021data} added disturbance to the interaction between the user and the project based on the probability generation model. Yue et al.~\cite{yue2021black} used distillation to extract the black-box model into a white-box model by synthesizing data and tags, and then use the white-box model to generate a black-box model against sample attacks. Chen et al.~\cite{chen2022knowledge} enhanced the generation of false user profiles through knowledge graphs, thereby increasing the effectiveness of attacks.

Another type of attack on recommender systems is the Data Poisoning Attack (DPA), also known as a shilling attack or injection attack, in which the attacker only knows the data but not the model information. The goal of this attack is to inject fake users and ratings into a recommender system to change the recommendation list. Although traditional algorithms, such as association rule-based algorithms and graph-based algorithms, have been used to solve the DPA problem, the majority of these attacks do not perform well. Tang et al.~\cite{tang2020revisiting} treated the DPA as an optimization problem and proposed adversarial gradient approximating methods. Huang et al.~\cite{huang2021data} simulated the target recommender system by training a surrogate model, but this problem remains unresolved as it is NP-hard to optimize the model without knowing it. Wu et al.~\cite{wu2021triple} proposed a triple adversarial learning model consisting of generator, discriminator, and influence modules and devise methods to estimate the influence of each generated fake user. Wu et al.~\cite{wu2022fedattack} proposed a poisoning method for federated recommendations, which subverts training and degrades model performance by false hardest samples.

There are two main categories: targeted and non-targeted. Targeted attacks, as the name suggests, have a predefined target class and are designed to mislead the recommender system from the ground truth class to a specific wrong class. For instance, Di et al.~\cite{di2020taamr} developed an attack model TAaMR to manipulate recommendations using FGSM and PGD. Meanwhile, Hu et al.~\cite{hu2019targeted} focused on recommending items to a specific group of users by injecting fake ratings and social connections in a factorization-based recommender system. These targeted attacks are driven by a clear motivation to influence recommendations for profits~\cite{tang2020revisiting}. Zhang et al.~\cite{zhang2022pipattack} designed a new attack method based on the popularity bias, which is achieved by increasing the exposure of the target (unpopular item).

On the other hand, non-targeted attacks do not have a specific target class and instead aim to attack the model to any class except the ground truth. Although non-targeted attacks have been understudied compared to targeted attacks, they are still an important aspect to consider in evaluating the robustness of recommender systems.

In conclusion, it is crucial to consider both targeted and non-targeted attacks in order to thoroughly assess the security and robustness of recommender systems. While targeted attacks have a strong motivation, non-targeted attacks are equally important to consider as they can also compromise the accuracy of recommendations.

To improve the robustness of the recommender system, defense models, such as adversarial training, and model distillation is often used. 
Generally, adversarial training is a min-max game between the adversary and the model, the adversary tries to maximize the likelihood of its attack success by searching for a proper perturbation $\sigma$, however, the model tries to minimize the risk by optimizing the model parameters $\theta$. The objective function is shown as follows:
\begin{equation}
\small
	\arg \min_{\theta} \max_{\sigma,\|\sigma\|\leq \epsilon} L(\theta+\sigma)
\end{equation}
where $\epsilon$ denotes the upper bound of perturbation. 
Due to the min-max game framework, some works try to categorize this type of defense method as a generative adversarial network, details please refer to~\cite{deldjoo2020recommender}.
As mentioned before, the perturbation can be added in different places in a recommender system.
Most of the works add it to the model parameters directly~\cite{anelli2021msap,yuan2019adversarial,christakopoulou2019adversarial}.
For example, Yuan et al.~\cite{yuan2019adversarial} added the perturbation to the weights of the encoder and decoder in their adversarial auto-encoder framework, then train the model with the perturbed weights to carry out adversarial training.
While, Anelli et al.~\cite{anelli2020empirical} added the perturbation to the image in the visual-based recommender system to replace the normal data, and conduct the defense with the adversarial training using the perturbed data directly. 
Distillation is another commonly used defense model, which is to transfer the knowledge from a teacher model to a student model by training a soft label, and the distilled network will be robust to the perturbed examples.
For example, Du et al.~\cite{du2018enhancing} designed the teacher-student framework over the neural collaborative filtering, in which knowledge is distilled successfully from the teacher model, and the robustness is validated under the C\&W attack. Yue et al.~\cite{yue2022defending} proposed a Dirichlet neighborhood sampling defense method for configuration file pollution. This method learns embedding through multi-hop neighbors and uses it for defense.
In conclusion, adversarial training and model distillation are two commonly used defense models to enhance the robustness of recommender systems. These methods address the challenge of robustness by adding perturbations and transferring knowledge, respectively.

\subsection{Data Bias}
Acquiring user behavior data for recommender systems through observation rather than experiments creates unavoidable deviations caused by various factors~\cite{chen2020bias}. Neglecting these deviations during RS design can lead to reduced model performance, affecting the user's experience and trust in the system. As a result, eliminating bias and improving the RS's robustness have emerged as crucial areas of research in the field. RS bias can arise from popularity bias, selection bias, exposure bias, and position bias. Addressing these biases is essential to ensure reliable and effective recommendations for users.

Popularity deviation can be understood as an item-based fairness problem (or a long-tail problem). It means that a very small number of items or items with relatively high popularity interact with users most of the time, while other items interact with users very little. Interactions here include clicks, bookmarks, inquiries, purchases, and post-purchase reviews. On the basis of this data, the results of model training will inevitably shift to high popularity, which is specifically manifested as giving higher scores to items or items with high popularity and giving lower scores to items or items with low popularity. Thus, the personalization level of RS is reduced, and the recommendation result is unfair. For this problem, this paper will explain the methods scholars deal with the popularity bias in Section ~\ref{sec:fairness}.

Selection bias refers to the deviation caused by the user's inability to completely rate each item. Specifically, users tend to rate the products they are interested in, very satisfied and very dissatisfied products, and the vast majority of products do not belong to these three categories, which causes the MissingNotAtRandom (MNAR) problem, resulting in deviation. Marlin et al.~\cite{marlin2012collaborative} provided researchers with reasonable evidence of selection bias in rating data. Steck et al.~\cite{steck2013evaluation} also proposed that the rating prediction is only related to the observed ratings. The distribution of the rating data is different from the distribution of all ratings, but the ranking usually considers all items in the collection, whether users rated them or not.

Wang et al.~\cite{wang2016learning} proposed a novel query-based bias estimation method, which can capture queries with similar results and handle sparse data. Wang et al.~\cite{wang2021combating} assumed that a small part of the training data is unbiased, and then used the unbiased data to adaptively assign the tendency weight to the biased training rating, which reduces the influence of selection bias on the recommendation system training, and reduces the influence of selection bias on recommendation system training through optimization algorithm. Liu et al.~\cite{liu2021self} proposed a self-supervised learning (SSL) framework Rating Distribution Calibration (RDC) to mitigate the negative impact of selection bias on RS and correct the estimated rating distribution by introducing a rating calibration loss. Huang et al.~\cite{huang2022different} proved through experiments that in dynamic scenarios where both selection bias and user preferences are dynamic, existing debiasing methods are no longer unbiased, so DebiAsing was introduced in dyNamiCscEnaRio (DANCER) to improve dynamic scoring prediction performance. Shi et al.~\cite{shi2023selection} proposed a temporal visibility-based data population strategy using uninteresting items to alleviate selection bias in the data, by using a weighted matrix factorization model to learn users' pre-use preferences for unrated items combined with user activities, item Popularity and temporal rating information are non-uniformly weighted to recover the true rating distribution of users.

Exposure bias means that the vast majority of users have limited time and can only see a part of the items exposed by the system and interact with these items. However, other unexposed items in the item library are unknown to the user, and there are also items that the user really likes or dislikes, so certain deviations will occur.

Zhou et al.~\cite{zhou2021contrastive} designed a CLRec model based on contrastive learning, which can improve the fairness, effectiveness, and efficiency of deep candidate generation (DCG) in RS with a large candidate size, and further improved the model to Multi-CLRec to accurately Reduce multi-intent perception bias and verify model performance through actual deployment on Taobao.
Wang et al.~\cite{wang2022unbiased} used the latent outcome framework to understand the bias mechanism, used inverse training instance reweighting to correct its propensity score (IPS), reformulated the sequential recommendation task, and further designed a parametric model to remove the influence of potential confounding factors, and passed Experiments have verified the effectiveness of the model.

Positional deviation means that users will basically choose items located in more prominent positions in the recommendation list (at the top, bold, etc.) for interaction, although these items are not optimally related to the user's needs, so a certain deviation occurs. Position deviation is crucial to the CTR (Predicting Click-Through Rate) problem in RS.

Collins et al.~\cite{collins2018position} conducted experiments on the digital book system Sowiport and the reference manager system JabRef, demonstrating the close relationship between the recommendation rating shown to the user and whether the user decides to click it. Agarwal et al.~\cite{agarwal2019estimating} estimated the user's propensity by obtaining specific types of intervention data from multiple different historical feedback logs to estimate the propensity and designed an extreme value estimator to further estimate the propensity. Guo et al.~\cite{guo2019pal} designed a Position-bias Aware Learning (PAL) framework to predict CTR (Click-Through Rate) in RS, the model can be trained offline and perform offline reasoning when there is no position information, the model has proved advanced in online experiments. 

This section contains a discussion of various forms of data bias, namely popularity bias, selection bias, exposure bias, and location bias. To mitigate these biases, researchers mainly rely on debiasing techniques, in addition to utilizing score calibration losses and propensity scores to improve model efficiency. These approaches aim to increase user confidence and trust in the system by ensuring data fairness and reducing discriminatory outcomes.

\subsection{Fairness}
\label{sec:fairness}
The fairness of the recommender system is proposed for a personalized recommendation. It can be divided into user-based fairness and project-based fairness. The former means that users should not be discriminated against by the recommendation system because of their own sensitive attributes, and the latter means that each item should have an equal chance to be recommended. Both the cold-start problem and the long-tail problem in the RS domain are typical cases of item-based fairness problems. It is worth noting that the long tail problem is also related to exposure bias.
\subsubsection{User-Based Fairness}
For user-based fairness, the current mainstream methods are mainly based on meta-learning, confrontation training, federated learning,  reinforcement learning, and Transformer, etc. 
For example, Finn et al.~\cite{finn2017model} proposed a meta-learning algorithm that can achieve better performance in small samples, which brings new ideas to solve the cold start problem and improve the fairness of RS. Lee et al.~\cite{lee2019melu} designed the MeLu meta-learning model to improve RS cold start ability and fairness and proposed an evidence candidate selection strategy to distinguish items. Dong et al.~\cite{dong2020mamo} designed two memory matrices to store task-specific memory and feature-specific memory, respectively guiding the initialization of the meta-learning model and predicting user preferences, improving cold-start performance. Dong et al.~\cite{dong2020mamo} designed two memory matrices to store task-specific memory and feature-specific memory, respectively guiding the initialization of the meta-learning model and predicting user preferences, improving cold-start performance.
Lu et al.~\cite{lu2020meta} proposed the MetaHIN model based on the idea of meta-learning, which improved the cold start ability of HIN. 
Wei et al.~\cite{wei2020fast} proposed the MetaCF model based on the CF model using the idea of meta-learning, this model can optimize the adaptive learning rate in a fine-grained manner and improve the cold start ability of CF. Feng et al.~\cite{feng2021cmml} designed a context-modulated meta-learning model, which is compatible with mainstream industrial deployments, reduces the computational efficiency of the model through feed-forward operations, and alleviates the cold-start problem. Wei et al.~\cite{wei2022comprehensive} proposed a comprehensive fair meta-learning framework called CLOVER, which comprehensively improves individual fairness, group fairness, and counterfactual fairness without degrading cold start performance. Neupane et al.~\cite{neupane2022dynamic} proposed a novel dynamic meta-learning model for historically active and currently inactive users, which can solve the time-sensitive cold start problem to a certain extent. Pang et al.~\cite{pang2022pnmta} proposed a pre-trained network modulation and task adaptation model (PNMTA) for user cold-start recommendation, the model is trained by a non-meta-learning method to obtain a pre-trained model, and the learned knowledge is transferred to the meta-learning phase modulation. The model takes better in the experiment to improve the cold start performance.

Apart from the meta-learning, Shivaswamy et al.~\cite{shivaswamy2022adversary} designed an adversarially trained machine learning model that improves service for under-served users. Liu et al.~\cite{liu2022fairness} communicated group statistical information during the federated optimization period and improved the fairness of the model while protecting privacy through differential privacy technology and privacy protection. Do et al.~\cite{do2022online} considered the envy problem of fairness auditing and regarded auditing as a new pure exploration problem in multi-armed bandit machines, proposed a sample-efficient algorithm, and analyzed the performance of the model. Li et al.~\cite{li2022transform} proposed the Cold-Transformer model, which transforms the cold state embedding into a warm state embedding by adaptively offsetting the feature distribution based on the context-based embedding, so as to express user preferences.

From the mentioned studies, it can be concluded that researchers are actively exploring different approaches to improve user-based fairness in machine learning models. Meta-learning, adversarial training, differential privacy, and multi-armed bandit algorithms are being used to address fairness issues and protect user privacy. Additionally, models like the Cold-Transformer are being developed to adaptively incorporate context-based embeddings to improve the accuracy of user preferences. Overall, these studies demonstrate the importance of addressing user-based fairness in machine learning models and the potential benefits of implementing these approaches to improve service for under-served users while maintaining user privacy.

\subsubsection{Item-Based Fairness}
To address the issue of item-based fairness, various methods have been proposed.  

One approach is based on causal inference, as proposed by Gupta et al.\cite{gupta2021causer}. They designed a comprehensive causal inference framework that can reduce popularity bias in the data generation and model training phases, and alleviate the long tail problem to some extent. Wei et al.\cite{wei2021model} used a causal graph to describe the causal relationship in the recommendation process and eliminated item popularity through counterfactual reasoning to enhance model fairness.

Another approach is based on adversarial training. Anelli et al.~\cite{anelli2021idiosyncratic} used adversarial regularization to overcome the vulnerability of the RS model to model weight interference and increase the model's accuracy. However, they found that in the case of unbalanced data distribution, adversarial regularization may amplify popularity deviation and reduce model fairness. Li et al.~\cite{li2022fairgan} proposed the FairGAN model based on a Generative Adversarial Network (GAN), which maps the fairness problem to the negative preference problem of implicit feedback and preserves user utility as much as possible while ensuring fairness.

Meta-learning and reinforcement learning are also used for item-based fairness. Du et al.~ \cite{du2019sequential} combined few-shot learning and meta-learning and designed the $S^2$Meta model to improve item-based fairness. Xie et al.~\cite{xie2022komen} proposed a meta-learning framework KoMen that first customizes global model parameters through scene-based domain knowledge and then learns scene-specific parameters through a hybrid expert architecture. This model can improve model cold-start performance. Ji et al.~\cite{ji2021reinforcement} proposed an actor-critic reinforcement learning (RL-LTV) framework that incorporates item lifetime value to improve cold-start performance and address the long-term return problem of RS systems. Recently, Shalaby et al.~\cite{shalaby2022m2trec} proposed the M2TRec model, which is a metadata-aware multi-task Transformer model based on session recommendation. This model integrates metadata to learn shared representations of different items and is able to recommend cold-start and sparse items.

In conclusion, various methods have been proposed for item-based fairness in RS, including causal inference, adversarial training, meta-learning, and reinforcement learning. Each method has its strengths and weaknesses, and the choice of method depends on the specific requirements of the RS system.

\section{Measurements}
Here are different kinds of evaluation measurements for the recommender system due to different tasks.
Generally, dues to the application type and prediction target, they can be categorized into two parts: Rating Based Indicator (RBI), and Item Based Indicator (IBI).
RBI means that the output of the recommender system is a rating score. A good recommendation can be inferred from a predicted rating that is close to the true rating.
IBI means the output of the recommender system is a set or a list of items. A good recommendation is based on the user's desired match between the predicted item and the target item.
In the sections below, we will give a detailed introduction to the indicators that are usually used in recent studies of these two types.

\subsection{Rating Based Indicator}
The rating-based indicator is to evaluate the quality of the prediction rating score. 
The direct way is to calculate the gap between implicit/explicit labels.

One most popular measurements are \textbf{R}oot \textbf{M}ean \textbf{S}quared \textbf{E}rror (RMSE) when the rating score is explicit value. The equation format for the RMSE can be expressed in Equation~\ref{equ:rmse}.
\begin{equation}
	\label{equ:rmse}
 \small
	RMSE = \sqrt{\frac{1}{|U||I|}\sum_{u\in U,i \in I}{(\hat{r}_{ui} - r_{ui})^{2}}}
\end{equation}
where $U$ is the set of the users, $I$ is the set of items, and $\hat{r}$ denotes the predicted rating, and $r$ denotes the true rating.
Same as RMSE, Mean Absolute Error (MAE) is another popular measurement that can be expressed in Equation~\ref{equ:mae}.
\begin{equation}
	\label{equ:mae}
 \small
	MAE = \sqrt{\frac{1}{\vert U\vert\vert I\vert}\sum_{u\in U,i \in I}{\vert \hat{r}_{ui} - r_{ui}\vert}}
\end{equation}
RMSE and MAE are non-negative, a lower RMSE and MAE are better than a higher one. While each value ($(\hat{r}_{ui}-r_{ui})^{2}$ in RMSE, $\vert\hat{r}_{ui}-r_{ui}\vert$ in MAE ) is proportional to the final error, which leads RMSE and MAE are sensitive to outliers.

\subsection{Item Based Indicator}
As the recommender system is a set or list of items, 
if there is no ranking information, the confusion matrix listed in Table~\ref{tab:confumatr} can be adopted in the evaluation.

\begin{table}[htp]
	\centering
		\caption{The confusion matrix for the recommender system.}
	\label{tab:confumatr}
 \tiny
	\begin{tabular}{lll}
		\toprule[0.8pt]
		& Recommended      & Not Recommended  \\\midrule[0.8pt]
		Used    & True Positive (TP)   & False Negative (FN)  \\
		Not Used & False Positive (FP)   & True Negative (TN)   \\\bottomrule[0.8pt]
	\end{tabular}

\end{table}

TP is the outcome of the used items that are recommended by the system, and FP is the outcome of not used items that aren't recommended by the system. FN denotes the outcome of the used items that are not recommended by the system, and TN is the outcome of the not used items that are not recommended by the system. Generally, a more comprehensive composition of those four values is adopted in the recommender system, such as $Precision$, $Recall$, and $F-Measure$ in Equation~\ref{equ:compre}.
\begin{equation}
	\label{equ:compre}
	\begin{aligned}
  \small
		\textit{Precision} &=\frac{TP}{TP+FP} \\
		\textit{Recall} &= \frac{TP}{TP+FN} \\
		\textit{F-Measure} & = \frac{2* \textit{Precision} * \textit{Recall}}{\textit{Precision} + \textit{Recall}} \\
	\end{aligned}
\end{equation}
\subsubsection{Ranking Indicators}
In fact, most recommendations are ordered lists of items, but $Precision$ and $Recall$ have nothing to do with order.
Therefore, it makes more sense to take top-$k$ $Precision$ and $Recall$ at cutoff $k$ ($Precision@k$ and $Recall@k$ for short), which considers only a subset of the ordered list from the 1st item to the $k$-th item. 

A more macro indicator is to average all the users to obtain the mean average precision ($MAP@k$) and mean average recall ($MAR@k$), which is depicted in Equ~\ref{equ:mapr}.
\begin{equation}
	\label{equ:mapr}
	\begin{aligned}
 \small
		MAP@k &= \frac{\sum_{i=1}^{k}(Precision@i \times rel(i))}{\min(k, \text{number of relevant items})} \\
		MAR@k &= \frac{\sum_{i=1}^{k}(Recall@i \times rel(i))}{\text{total number of relevant items}}
	\end{aligned}
\end{equation}
where $rel(i)$ is an indicator function, if the $i$-th item is relevant to the target, it is equal to 1, otherwise is 0.
With these two macro indicators, we can find out the general performance of the recommender system in the real application.
However, these two indicators have some drawbacks, for example, 
they are not fit for fine-grained numerical ratings.
They require a threshold to convert fine-grained numerical ratings to binary correlations, which can lead to bias and loss of fine-grained information.

To solve the problem that exists in $MAP@k$ and $MAR@k$, and take the ranking into the consideration,
another widely used metric is normalized discounted cumulative gain (NDCG)~\cite{wang2013theoretical}. And it has become a popular evaluation metric in ranking-based recommender systems.
Like $MAP@k$ and $MAR@k$, the goal of $NDCG$ is to put highly relevant items at the top of the recommendation list, but it improves the recommendation evaluation quality with fine-grained numerical ratings.
$NDCG$ is inferred from the standard discounted cumulative gain (DCG) which is defined as $DCG = \sum_{i}^{p}\frac{rel(i)}{\log_{2}(i+1)}$, where $rel(i)$ is the fine-grained numerical ratings. There is a logarithmic operation that penalizes the relevant score in proportion to the location of the item. Based on $DCG$, we need to obtain the maximum possible $DCG$ at position $p$ to get the ideal DCG (IDCG for short), and this is defined as $IDCG =\sum_{i=1}^{|rel(p)|} \frac{2^{rel(i)}-1}{\log_{2}(i+1)}$, where $|rel(p)|$ is the number of relevant items up to position $p$. $IDCG$ provides a nice normalization factor for $NDCG$ which is defined in Equ~\ref{equ:ndcg}.
\begin{equation}
	\label{equ:ndcg}
 \small
	NDCG = \frac{DCG}{IDCG}
\end{equation}
However, there are some disadvantages of using $NDCG$, for example, it cannot handle incomplete ratings, in this case, the missing value can be filled with mean/median values before it is passed to $NDCG$. 

If there is no user's relative ranking information for two items, then we can tie them together into a single item. In the evaluation metrics mentioned above, except $NDCG$, there is no such relative information. Another metric that is suitable for such a situation is the normalized distance-based performance measurement (NDPM). Given reference ranking $r_{ui}$ and system rankings $\hat{r}_{ui}$ of $n_{u}$ items $i$ for user $u$, the definition of NDPM is given by Eq.~\ref{equ:ndpm}. 

\begin{equation}
	\label{equ:ndpm}
	\begin{aligned}
 \small
		C^{+} &=\sum_{ij}\text{sgn} (r_{ui}-r_{uj})\text{sgn} (\hat{r}_{ui}-\hat{r}_{uj})\\
		C^{-} &=\sum_{ij}\text{sgn} (r_{ui}-r_{uj})\text{sgn} (\hat{r}_{uj}-\hat{r}_{ui})\\
		C^{u}& = \sum_{ij}\text{sgn}^{2}(r_{ui}-r_{uj})\\
		% C^{s} = \sum_{ij} \text{sgn}^{2}(\hat{r}_{ui}-\hat{r}_{uj})\\
		C^{t} &= C^{u} - (C^{+}+C^{-}) \\
		NDPM &= \frac{C^{-}+0.5C^{t}}{C^{u}} \\
	\end{aligned}
\end{equation}

where $C^{u}$ is the number of pairs in the ordered list (i.e., not tied), $C^{+}, C^{-}$ are the number of the pairs in the correct order and in the incorrect order respectively, $C^{t}$ is the number of pairs where the reference ranking does not tie but the system ranking ties. 
NDPM focuses on predictive rankings rather than specific predictive ratings, so it is not applicable to systems designed to provide users with accurate predictive ratings. 

\begin{table}[htp]
	\centering % centering table
		\caption{The similarity methods used in memory-based CF. In the table $r_{u,i}$ denotes the rating that user \(u\) gives to items
		\(i\), \(r_{med}\) Indicates the median value in the rating scale (e.g.
		3 in the rating scale of 5), $\hat{r}_{u}$ is the average rating of user $u$ gives to the co-rated items with $v$, \(\hat{\dot{r}}_{u}\) is the average rating of user $u$ gives to all the items, \(I\) is the set of all items, if there is no rate for user \(u\) on a item in \(i\in I\), the rating \(r_{u,i}\) is zero, $I_{u,v}$ indicates the co-rated items by user $u$ and $v$,\(\vert I_{u}\cup I_{v}\vert\) is the total item number rated by user \(u\) and \(v\), $\hat{r}_{i}$ is the average rating of item $i$, \(H\) is the experimental value based on the performance, details about $ Proximity(r_{u,i},r_{v,i}), Impact(r_{u,i},r_{v,i}), Popularity(r_{u,i},r_{v,i})$ please refer to~\cite{ahn2008new}, $\sigma_{u}$ represents the standard variance of user $u$ .} 
	\label{tab:sim}
 \tiny
	\begin{tabular}{m{0.15\columnwidth}m{0.75\columnwidth}}
		\toprule[0.8pt]
		\textbf{Measurements} & \textbf{Definitions}\tabularnewline \midrule[0.8pt]
		Pearson Correlation Coefficient (PCC) &
		$S^{PCC}_{u,v} = \frac{\sum_{i\in I_{u,v}}(r_{u,i}-\hat{r}_{u})(r_{v,i} - \hat{r}_{v})}{\sqrt{\sum_{i\in I_{u,v}}(r_{u,i}-\hat{r}_{u})^{2}}\sqrt{\sum_{i\in I_{u,v}}(r_{v,i}-\hat{r}_{v})^{2}}}$ \tabularnewline \midrule[0.35pt]
		Weighted Pearson Correlation Coefficient (WPCC) &
		\(S^{WPCC}_{u,v}=\begin{cases} S^{PCC}_{u,v}\cdot \frac{\vert  I_{u,v}\vert}{H}, & \text{ if }\vert I_{u,v}\vert\leq H \\ S^{PCC}_{u,v}, & \text{ otherwise } \end{cases}\) \tabularnewline\midrule[0.35pt]
		Sigmoid-based Pearson Correlation Coefficient (SPCC) &
		\(S^{SPCC}_{u,v} = S^{PCC}_{u,v}\cdot \frac{1}{1+\text{exp}(-\vert I_{u,v}\vert/2)}\) \tabularnewline\midrule[0.35pt]
		Cosine Similarity (COS) &
		\(S^{COS}_{u,v} = \frac{\sum_{i\in I_{u,v}}r_{u,i}r_{v,i}}{\sqrt{\sum_{i\in I_{u,v}}r_{u,i}^{2}}\sqrt{\sum_{i\in I_{u,v}}r_{v,i}^{2}}}\) \tabularnewline \midrule[0.35pt]
		Adjusted Cosine (ACOS) &
		\(S^{ACOS}_{u,v} = \frac{\sum_{i\in I}(r_{u,i}-\hat{\dot{r}}_{u})(r_{v,i} - \hat{\dot{r}}_{v})}{\sqrt{\sum_{i\in I}(r_{u,i}-\hat{\dot{r}}_{u})^{2}}\sqrt{\sum_{i\in I}(r_{v,i}-\hat{\dot{r}}_{v})}}\)\tabularnewline \midrule[0.35pt]
		Constrained Pearson's Correlation Coefficient (CPCC) &
		\(S^{PCC}_{u,v} = \frac{\sum_{i\in I_{u,v}}(r_{u,i}-r_{med})(r_{v,i} - r_{med})}{\sqrt{\sum_{i\in I_{u,v}}(r_{u,i}-r_{med})^{2}}\sqrt{\sum_{i\in I_{u,v}}(r_{v,i}-r_{med})^{2}}}\) \tabularnewline\midrule[0.35pt]
		Jaccard &
		\(S^{Jaccard}_{u,v} = \frac{\vert I_{u,v}\vert}{\vert I_{u} \cup I_{v}\vert}\)\tabularnewline\midrule[0.35pt]
		Mean Squared Difference (MSD) &
		\(S^{MSD}_{u,v}=1-\frac{\sum_{i\in I}(r_{u,i}-r_{v,i})^{2}}{\vert I_{u,v}\vert}\) \tabularnewline\midrule[0.35pt]
		Proximity Impact Popularity (PIP) &
		\(S^{PIP}_{u,v} = Proximity(r_{u,i},r_{v,i}) \cdot Impact(r_{u,i},r_{v,i}) \cdot \)  \(Popularity(r_{u,i},r_{v,i})\)\tabularnewline\midrule[0.35pt]
		%Proximity Significance Singularity (PSS) & $S^{PSS}_{u,v} = \sum_{i\in I_{u,v}}Proximity(r_{u,i},r_{v,i})\cdot Significance(r_{u,i},r_{v,i})\cdot Singularity(r_{u,i},r_{v,i})$\tabularnewline \hline
		Proximity Significance Singularity (PSS) & $S^{PSS}_{u,v} = \sum_{i\in I_{u,v}}(1-\frac{1}{1+\exp(-\vert r_{u,i}-r_{v,i}\vert)})\cdot (\frac{1}{1+\exp(-\vert r_{u,i}-r_{med}\vert\cdot \vert r_{v,i}-r_{med}\vert)})\cdot (1-\frac{1}{1+\exp(-\vert (r_{u,i}+r_{v,i})/2-\hat{r}_{i}\vert})$\tabularnewline \midrule[0.35pt]
		Jaccard Proximity Significance Singularity (JPSS) & $S^{JPSS}_{u,v} = S^{PSS}_{u,v}\cdot S^{Jaccard}_{u,v} $\tabularnewline \hline
		User Rating Preference (URP) & $S^{URP}_{u,v} = 1-\frac{1}{1+\exp(-\vert \hat{r}_{u}-\hat{r}_{v}\vert\cdot \vert\sigma_{u}-\sigma_{v}\vert)}$\tabularnewline \midrule[0.35pt]
		New Heuristic Similarity Model (NHSM)& $S^{NHSM}_{u,v} = S^{JPSS}_{u,v}\cdot S^{URP}_{u,v} $\tabularnewline 
		\bottomrule[0.8pt]
	\end{tabular}
\end{table}

\subsubsection{Similarity}
\label{subsec:sim}
There are different ways to calculate the similarity that is proposed in recent years, such as Pearson correlation coefficient (PCC), cosine similarity (COS), mean squared difference (MSD), proximity impact popularity (PIP), Jaccard, proximity significance singularity (PSS), new heuristic similarity model (NHSM) and etc. Details are listed in Table~\ref{tab:sim}.

In PCC, $r_{u,i}$ denotes the rating that user $u$ gives to items $i$. $\hat{r}_{u}, \hat{r}_{v}$ are taken over their co-rated items only. The range of the similarity is $[-1, 1]$, if $S_{PCC}=0$, it means there is no correlation between these two users, $S_{PCC}=1$ indicates the positive correlation, $S_{PCC}=-1$ is the negative correlation.
The computation complexity and accuracy depend on the number of co-rated items. Typically, the more co-rated items, the higher the accuracy and computation complexity at the same time. 
Khawar et al.~\cite{khawar2019cleaned} argue that, unlike the PCC, the COS naturally possesses the desirable property of eigenvalue shrinkage for large eigenvalues.
In the equation of ACOS, $I$ is the set of all items, if there is no rate for user $u$ on an item in $i\in I$, the rating $r_{u,i}$ is zero. Different from PCC, $\hat{r}_{u}$ and $\hat{r}_{v}$ are calculated from all ratings of user $u$ and $v$.

\section{Conclusion}
In conclusion, this survey paper provides a comprehensive overview of the recent developments in recommender systems. The field has seen significant progress in the past few years, and it is evident from the various innovative approaches and algorithms that have been proposed. The diversity of recommender systems and their applications, combined with the growing amount of data and computing power, has created a rich and dynamic environment for research. According to a recent article collection from Applied Soft Computing~\cite{malandri2022soft}, Recommender systems and Sentiment Analysis are two crucial applications of algorithms, and the authors have conducted a bibliometric analysis of research articles using structural topic modeling. Their analysis shows that in the past two decades, research focused on soft computing has shifted from recommender systems to sentiment analysis. There is expected to research in this area, such as using fuzzy logic for aspect-based sentiment analysis or sentiment knowledge base reasoning.

Our discussion started with an overview of the current trends in personalized and group-based recommender systems. For personalized recommender systems, we went beyond the conventional categorization by including knowledge-based systems which have seen a surge in popularity due to the recent improvements in deep learning methods. The personalized recommender system was then divided into four distinct categories: collaborative filtering, content-based, knowledge-based, and hybrid.

Furthermore, our survey delved into the idea of group-based recommender systems, which can be categorized into memory-based and model-based techniques.In addition, we also discussed item-oriented group recommendation, bundled recommendation. To give a complete overview of the field, we also discussed the common challenges faced by recommender systems, such as robustness issues ,data bias and fairness issues. Lastly, we analyzed various performance metrics, including those based on ratings, ranking, and similarity. Overall, this survey aimed to provide a comprehensive and up-to-date summary of recommender systems to provide a better understanding of the current landscape of the field.

The rich and dynamic research environment of recommender systems has been created by the diversity of its applications and the growing amount of data and computing resources available. Although the field still faces challenges, such as robustness ,data bias and fairness issues, the future of recommender systems holds great promise and the potential to greatly benefit the human race.

Currently, with the prevalence of ChatGPT, which is based on Generative Pre-Training (GPT) technology, researchers are considering its potential to enhance their respective fields, including recommender systems (RS). Within the RS domain, conversational recommendation systems that generate user recommendations through dialogue are gaining popularity. For instance, Gao et al. \cite{gao2023chat} improved the interaction and interpretability of the recommendation system by leveraging ChatGPT and also enabled cross-domain recommendation capabilities. Additionally, Cui et al. \cite{cui2022m6} proposed a recommendation system based on M6, which is similar to ChatGPT and T5. It is expected that GPT technology will catalyze novel advances in the field of recommender systems.

\bibliographystyle{IEEEtran}
\bibliography{references}

\end{document}